\documentclass[11pt]{article}
\usepackage{titlesec}
\usepackage[utf8]{inputenc}
\usepackage{amsmath}
\usepackage{graphicx}
\usepackage{braket}
\usepackage{amsfonts}
\usepackage{amssymb}
\usepackage{geometry}
\usepackage{xfrac} 
\usepackage{physics} 
\usepackage{enumerate}
\usepackage{setspace}
\usepackage{multicol} 
\usepackage{xcolor} 
\usepackage{float}
\usepackage[nottoc,numbib]{tocbibind}
\usepackage{multirow} 
\usepackage{lscape}
\usepackage{diagbox}
\usepackage{hyperref}
\usepackage{subfig}
\usepackage{yfonts}
\usepackage{mathrsfs}
\usepackage{ulem}
\usepackage{tabularx} 
\usepackage{bm}
\usepackage{mathtools}
\usepackage{amsbsy}
\usepackage{dsfont}
\usepackage[scr=bickham]{mathalfa}
\usepackage{lscape}
\usepackage{pdflscape}

\usepackage[mathscr]{euscript} 
\usepackage{longtable}

\newcommand{\paren}[1]{\left( #1 \right)}
\newcommand{\loga}[1]{ \log{\left(#1\right)}}

\newcommand{\corche}[1]{\left[#1\right]}
\newcommand{\llave}[1]{ \left \{ #1\right \}}

\newcommand{\M}{M}
\newcommand{\dif}{\text{d}}
\newcommand{\dx}{\text{d}x}
\newcommand{\dt}{\text{d}t}
\newcommand{\du}{\text{d}u}

\newcommand{\expo}[1]{\mathrm{e}^{#1}}
\newcommand{\pe}{\alpha}

\newcommand{\On}{\Omega_{n}^\lambda}

\newcommand{\x}{\hat{x}}
\newcommand{\Ona}{\Omega_{0}^\lambda}
\newcommand{\Onb}{\Omega_{1}^\lambda}
\newcommand{\Onc}{\Omega_{2}^\lambda}
\newcommand{\Ond}{\Omega_{3}^\lambda}

\newcommand{\G}[1]{\Gamma\left(#1\right)}

\newgeometry{bottom=2.5cm, top=2.5 cm, left=2.5cm, right=2.5cm}

\title{Rényi and Tsallis information entropies for the Darboux III quantum nonlinear oscillator 
}

\begin{document}






\maketitle

\begin{center}

{\sc Ignacio Baena-Jimenez$^{1}$, Angel Ballesteros$^{1}$, Ivan Gutierrez-Sagredo$^{2}$ and Javier Relancio$^{2,3}$}

\medskip
{$^1$Departamento de F\'isica, Universidad de Burgos, 
09001 Burgos, Spain}

{$^2$Departamento de Matem\'aticas y Computaci\'on, Universidad de Burgos, 
09001 Burgos, Spain}

{$^3$ Departamento de
Física Teórica, Universidad de Zaragoza, 50009 Zaragoza,
Spain}
\medskip
 
e-mail: {\href{mailto:ibaena@ubu.es}{ibaena@ubu.es}, \href{mailto:angelb@ubu.es}{angelb@ubu.es}, \href{mailto:igsagredo@ubu.es}{igsagredo@ubu.es}, \href{mailto:jjrelancio@ubu.es}{jjrelancio@ubu.es}}

\end{center}

\begin{abstract}
 
The Darboux III oscillator is an exactly solvable $N$-dimensional nonlinear oscillator defined on a radially symmetric space with non-constant negative curvature. Its one-dimensional version can be seen as a position dependent mass system whose mass function $\mu = \paren{1 + \lambda x^2}$ depends on the nonlinearity parameter $\lambda$, such that in the limit $\lambda \to 0$ the harmonic oscillator is recovered. In this paper, a detailed study of the entropic moments and of the Rényi and Tsallis information entropies for the quantum version of the one-dimensional Darboux III oscillator is presented. In particular, analytical expressions for the aforementioned quantities in position space are obtained. Since the Fourier transform of the Darboux III wave functions does not admit a closed form expression, a numerical analysis of these quantities has been performed. Throughout the paper the interplay between the entropy parameter $\pe$ and the nonlinearity parameter $\lambda$ is analysed, and known results for the Shannon entropy of the Darboux III and for the Rényi and Tsallis entropies of the harmonic oscillator are recovered in the limits $\pe \to 1$ and $\lambda \to 0$, respectively. Finally, motivated by the strong non-linear effects arising when large values of $\lambda$ and/or highly excited states are considered, an approximation to the probability density function valid in those regimes is presented. From it, an analytical approximation to the probability density in momentum space can be obtained, and some of the previously observed effects arising from the interplay between $\pe$ and $\lambda$ can be explained.

\end{abstract}

\bigskip




\renewcommand{\listfigurename}{LISTA DE FIGURAS}
\renewcommand{\listtablename}{Lista de Tablas}
\renewcommand{\figurename}{Figure}
\renewcommand{\tablename}{Table} 


%


\tableofcontents{}

\section{Introduction}


Position-dependent mass (PDM) systems have found applications across a wide range of physical contexts. 
Such models, that in general present highly nonlinear dynamics, have been applied, for instance, to the modeling of graded graphene vibrations \cite{dong2018nonlinear} and the analysis of vibrational resonance phenomena \cite{roy2021vibrational}. Also, in gravitational physics, PDM models have been studied in cosmological settings \cite{morris2015new,ballesteros2017hamiltonians,mustafa2023pdm,ahmed2025pdm} and in the context of quantum gravity \cite{lawson2022minimal,westphal2021measurement}. In solid-state physics, PDM formulations have been considered in relation to the theory of semiconductors and their relation to optical and electronic properties \cite{el2020generalized,sarker2017position}. These systems have also been applied to the study of optical properties of quantum wells and quantum dots, where spatially varying effective mass can significantly alter confinement and transition dynamics \cite{zhao2020influence,ghosh2016influence,kasapoglu2021effects}.

A particularly interesting example of a PDM system is given by the one-dimensional version the so called Darboux III nonlinear oscillator \cite{ballesteros2008maximally,ballesteros2011new,ballesteros2011quantum}, a $N$-dimensional maximally superintegrable  Hamiltonian representing an oscillator defined on a radially symmetric space with non-constant negative curvature, this corresponds to the $N$-dimensional generalisation of the so-called Darboux surface of type III\cite{darboux1915leccons}. In this paper we will deal with such one-dimensional version, which can be interpreted as a PDM system over the real line with a position-dependent mass $\mu = \paren{1 + \lambda x^2}$, a type of mass function suitable to describe certain semiconductor heterostructures \cite{SCHMIDT2006459,Koc2005}. The exact solvability of the quantum version of this model implies that analytical expressions for its eigenvalues and eigenfunctions are known, which makes it possible to get a very complete knowledge of many properties of this nonlinear $\lambda$-deformation of the harmonic oscillator model.

The aim of this paper is to present a complete study of the entropy-based measures for the one-dimensional Darboux III system, since only its Shannon entropy has been analysed in \cite{ballesteros2023shannon}. As it is well known, these quantities provide valuable insights into the localization, spread and complexity of the corresponding wavefunctions \cite{lopez2015statistical,hall1999universal,bialynicki2006formulation,zozor2007classes,puertas2017heisenberg}. Then we will obtain full analytical results for density probabilities in position space, together with numerical and  approximate analytical results in momentum space. In fact,  some novel features of the Darboux III nonlinear oscillator for high values of $\lambda$ and/or its quantum number $n$ will arise as a byproduct of the approach here presented.

We recall that Shannon entropy \cite{shannon1948mathematical} is one of the most widely used measures of information and has become a cornerstone of numerous applications in information theory \cite{thomas2006elements,gray2011entropy}. Rényi \cite{renyi1961measures} and Tsallis \cite{tsallis1988possible} entropies have been introduced as generalisations of the Shannon entropies depending on a parameter $\pe$. The properties and applications of these one-parameter generalisations of the Shannon entropy have also been thoroughly investigated (see, for instance, \cite{aczel1975measures,jizba2004world,leonenko2008class,sen2011statistical,jizba2015role,olendski2020renyi} and references therein).

Rényi and Tsallis entropies of the probability density $\rho(z)=\abs{\Psi(z)}^2$, which characterizes the quantum state $\Psi(z)$ of a one-dimensional system, are defined, respectively, as
\begin{align} \label{eq: renyi entropy}
    \mathcal{R}^{(\pe)} \left[ \rho\right]&=\frac{1}{1-\pe} \loga{ \mathcal{W}^{(\pe)} \corche{\rho}}, \hspace{1cm}  \pe > 0, \ \pe \neq 1, 
 \end{align}   
 \begin{align} \label{eq: tsallis entropy}
    \mathcal{T}^{(\pe)} \left[ \rho\right]&=\frac{1}{1-\pe} \paren{ \mathcal{W}^{(\pe)} \corche{\rho}-1}, \hspace{1cm}  \pe > 0, \ \pe \neq 1,   
\end{align}
where 
the symbol $\mathcal{W}^{(\pe)} \corche{\rho}$ denotes the frequency or entropic moment of order $\pe$ \cite{romera2001hausdorff,jizba2016one} of the density given by
\begin{align}
     \mathcal{W}^{(\pe)} \left[\rho \right] &=\int_{-\infty}^{+\infty} \rho^\pe(z)\dif z.
\end{align}
When $\pe \to 1$, the limit of Rényi (\ref{eq: renyi entropy}) and Tsallis (\ref{eq: tsallis entropy}) entropies converges to Shannon entropy:
\begin{align} \label{eq: shannon}
    \mathcal{S}\left[ \rho\right]=\lim_{\pe \to 1} \mathcal{R}^{(\pe)} \left[ \rho\right]=\lim_{\pe \to 1} \mathcal{T}^{(\pe)} \left[ \rho\right]=- \int_{\mathbb{R}}  \rho(z) \loga{\rho(z)} \dif z.
\end{align}

Shannon, Rényi, and Tsallis entropies quantify the spread of a probability distribution, but each one of them emphasises different features. Shannon entropy is additive and widely applicable when a standard measure suffices, thus playing an essential role in classical information and communication  \cite{shannon1948mathematical}. Rényi entropy, also additive, introduces a tunable parameter $\alpha$ that adjusts sensitivity to different regions of the distribution. In particular, when $\pe>1$, events with high probability are emphasised within the computation, while low probability ones become relevant when $\pe<1$. This makes it useful in contexts like multifractal analysis and systems with power-law behaviour \cite{yujun2017multiscale,chen2012renyi,bashkirov2000information}. Tsallis entropy is a non-additive generalisation with the same kind of  tunable parameter $\pe$, which is especially suited for systems with long-range correlations, memory effects, or heavy-tailed distributions \cite{tsallis2009nonadditive}. Rather than being interchangeable, these entropies are best understood as complementary tools, each appropriate under different physical or statistical conditions.


As a generic property, we have that
\begin{align} \label{eq: p effect}
    \mathcal{R}^{(\pe)} \corche{\rho} \geq
    \mathcal{R}^{(\beta)} \corche{\rho}, \hspace{1cm}     \mathcal{T}^{(\pe)} \corche{\rho} \geq
    \mathcal{T}^{(\beta)} \corche{\rho},   \hspace{1cm} \mathrm{if} \ \pe \leq \beta.
\end{align}
For continuous distributions, the Rényi entropy is not bounded from above. The Tsallis entropy also diverges in the limit of a maximally spread-out density when $\pe < 1$. However, for $\pe > 1$, the Tsallis entropy is bounded from above, and the larger $\pe >1$ is, the faster the Tsallis entropy approaches saturation. This provides a measure, controlled by $\alpha$, of how close the probability density is to a constant function.

Given the relevance of entropic measures in characterising quantum systems, we also consider the entropy-based uncertainty principle \cite{bialynicki2006formulation}. The Sobolev inequality for conjugated Fourier transforms  \cite{olendski2019renyi} in one dimension is given by:
\begin{align} \label{eq: incertidumbre sobolev}
    \paren{\frac{\pe}{\pi}}^{\frac{1}{4\pe}} \paren{\int_{\mathbb{R}} \rho^\pe(x)  \dif x}^{\frac{1}{2\pe}} \geq \paren{\frac{\beta}{\pi}}^{\frac{1}{4\beta}} \paren{\int_{\mathbb{R}}  \gamma^\beta(p) \dif p}^{\frac{1}{2\beta}},\hspace{1cm} \frac{1}{\pe}+\frac{1}{\beta} = 2, \hspace{0.7cm} \frac{1}{2} <  \pe \leq 1,
\end{align}
where $\rho(x)$ and $\gamma(p)$ are the probability densities in position and momentum spaces, respectively.

Applying this principle (\ref{eq: incertidumbre sobolev}) to the definition of Rényi entropy  (\ref{eq: renyi entropy}) yields:
\begin{align}
    \mathcal{R}^{(\pe)} \left[ \rho(x)\right]+ \mathcal{R}^{(\beta)} \left[ \gamma(p)\right] \geq
       -\frac{1}{2} \paren{\frac{1}{1-\pe} \log{\frac{\pe}{\pi}}+\frac{1}{1-\beta} \log{\frac{\beta}{\pi}}},\hspace{1cm} \frac{1}{\pe}+\frac{1}{\beta} = 2,
\end{align}
which is the Bialynicki-Birula formulation \cite{bialynicki2006formulation} of the uncertainty principle. This can be easily rewritten using  properties of the logarithm to get the Rényi-entropy-based uncertainty relation of Zozor-Portesi-Vignat \cite{zozor2008some} in one dimension
\begin{align} \label{eq: zozor} 
       \mathcal{R}^{(\pe)} \left[ \rho(x)\right]+ \mathcal{R}^{(\beta)} \left[ \gamma(p)\right] \geq
       \log{\left( \pi \pe^\frac{1}{2\pe-2} \beta^{\frac{1}{2\beta-2}}  \right)}, \hspace{1cm} \frac{1}{\pe}+\frac{1}{\beta} = 2.
\end{align}
Tsallis entropy-based uncertainty relation \cite{rajagopal1995sobolev} is derived by rewritting the Sobolev inequality (\ref{eq: incertidumbre sobolev}) in terms of the Tsallis entropy definition (\ref{eq: tsallis entropy}) with the same constraints, $\frac{1}{\pe}+\frac{1}{\beta} = 2$ and $\frac{1}{2} <  \pe \leq 1$,
\begin{align} \label{eq: uncertainty Tsallis}
    \paren{\frac{\pe}{\pi}}^{\frac{1}{4\pe}} \paren{\paren{1-\pe}\mathcal{T}^{(\pe)}\corche{\rho(x)}+1}^\frac{1}{2\pe} \geq \paren{\frac{\beta}{\pi}}^{\frac{1}{4\beta}} \paren{\paren{1-\beta}\mathcal{T}^{(\beta)}\corche{\gamma(p)}+1}^\frac{1}{2 \beta}.
\end{align}

These information measures for nonlinear oscillators \cite{olendski2023one} and PDM systems have been a subject of interest. For example, the Shannon entropy has been computed for a PDM system with a hyperbolic well \cite{guo2015shannon} or for the Mathews–Lakshmanan oscillator \cite{da2025kappa}. The Rényi entropy has also been used to study PDM systems, for instance in the context of the Frost–Musulin potential \cite{idiodi2016entropy}. It has also  been employed to analyze systems influenced by curvature, such as multifractal structures \cite{esquivel2017multifractal} or entanglement in curved geometries \cite{lee2014renyi}.

In the case of the Darboux III model, Shannon entropies were computed in \cite{ballesteros2023shannon} as functions of the excitation level $n$ and the parameter $\lambda$, which controls the position-dependent mass function. For $\lambda = 0$, the mass becomes constant and the system reduces to the harmonic oscillator (as will be shown in Section~\ref{sec:1D-DarbouxIII}). Shannon entropy was found to increase with both $\lambda$ and $n$ in position space, while it decreases with the same parameters in momentum space. We stress that this analysis focused on low-lying excited states and small values of $\lambda$. It is therefore natural to analyse whether the rest of entropic measures can be computed for the Darboux III nonlinear oscillator. Moreover, higher excited states and larger values of the parameter $\lambda$ are worth to be explored as well. We recall that the Shannon and Rényi entropies for the harmonic oscillator have been analytically computed in \cite{dehesa2019shannon,puertas2018exact}, and these results should be recovered in the limit $\lambda\to 0$.


Finally, we outline here the structure of the paper. In Section~\ref{sec:1D-DarbouxIII}, we introduce the one-dimensional Darboux III oscillator, recall its main properties, and fix the notation used throughout the paper. Section~\ref{sec:position} presents analytical expressions for the entropic moments in position space, which generalise those of the harmonic oscillator, and allow the computation of Rényi and Tsallis entropies for arbitrary $n$ and $\lambda$. In Section~\ref{sec:momentum}, we focus on momentum space, where the lack of analytical wave functions leads us to a numerical approach. Furthermore, we show that entropic uncertainty relations are always satisfied, and we analyze the dependence of the entropic results in terms of the nonlinearity parameter $\lambda$. In Section~\ref{sec: large lambda}, we explore the regime of highly excited states for sufficiently large values of $\lambda$. This provides further insight into the behaviour of Rényi and Tsallis entropies for this PDM system, and to the structure of the Darboux III wavefunctions. Finally, a concluding Section \ref{sec: conclusiones} closes the paper.

\section{The one-dimensional Darboux III oscillator} 
\label{sec:1D-DarbouxIII}

In this Section we recall the definition of the one-dimensional Darboux III nonlinear oscillator, both classical and quantum, and then we present the energy spectrum and eigenfunctions of the latter. 

The classical $N-$dimensional Darboux III oscillator is an exactly solvable maximally superintegrable \cite{ballesteros2008maximally} model representing a $N-$dimensional nonlinear oscillator defined on a radially symmetric space with nonconstant negative curvature. Its Hamiltonian is given by
\begin{align} \label{eq: Hn}
    H^{(N)}(\boldsymbol{x},\boldsymbol{p})=T(\boldsymbol{x},\boldsymbol{p})+V(\boldsymbol{x})=\frac{\boldsymbol{p}^2}{2 (1+\lambda \boldsymbol{x}^2)}+\frac{\omega^2 \boldsymbol{x}^2}{2 (1+\lambda \boldsymbol{x}^2)},
\end{align}
where parameters $\lambda, \omega > 0$ are real numbers and $(\boldsymbol{x},\boldsymbol{p}) \in \mathbb{R}^{2N}$ are canonical coordinates. In the limit $\lambda \to 0$, the hamornic oscillator Hamiltonian is recovered
\begin{align}
    H_0^{(N)}(\boldsymbol{x},\boldsymbol{p})=\lim_{\lambda \to 0}  H_0^{(N)}(\boldsymbol{x},\boldsymbol{p})=T_0(\boldsymbol{p})+V_0(\boldsymbol{x})=\frac{\boldsymbol{p}^2}{2}+\frac{\omega^2 \boldsymbol{x}^2}{2}.
\end{align}

In one dimension, the Hamiltonian (\ref{eq: Hn}) reduces to
\begin{align} \label{eq: H class}
H(x,p)=\frac{p^2}{2 (1+\lambda x^2)}+\frac{\omega^2 x^2}{2 (1+\lambda x^2)},
\end{align}
where the canonical coordinates are now $(x,p) \in \mathbb{R}^{2}$. 
The Darboux III oscillator can be thus interpreted as a deformation of the one-dimensional harmonic oscillator, as its Hamiltonian reduces to that of the harmonic oscillator in the limit $\lambda \to 0$, 
\begin{align} \label{eq: H0 class}
\mathcal{H}_0(x,p)=\mathcal{T}_0(x,p)+\mathcal{U}_0(x)=\frac{p^2}{2}+\frac{\M \omega^2 x^2}{2}.
\end{align}
Although there is no meaningful notion of spatial curvature in a one-dimensional setting, this system can still be interpreted as a nonlinear  oscillator with a position-dependent mass $\mu = \paren{1 + \lambda x^2}$.

\begin{figure}[H]
\begin{tabular}{cccc}
\subfloat[]{\includegraphics[scale = 0.9]{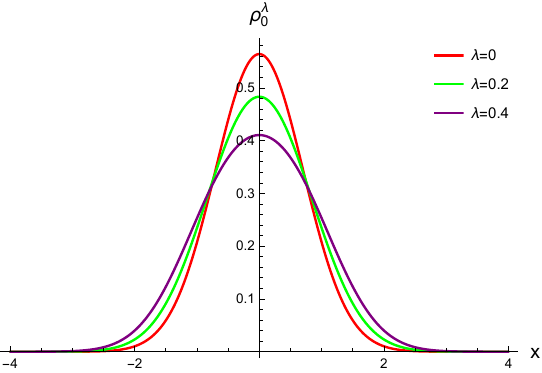}} &
\subfloat[]{\includegraphics[scale = 0.9]{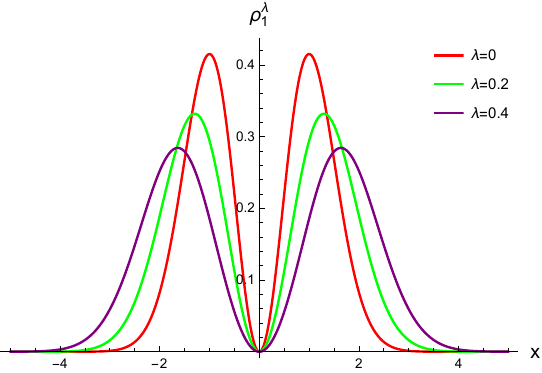}}  \\
\subfloat[]{\includegraphics[scale = 0.9]{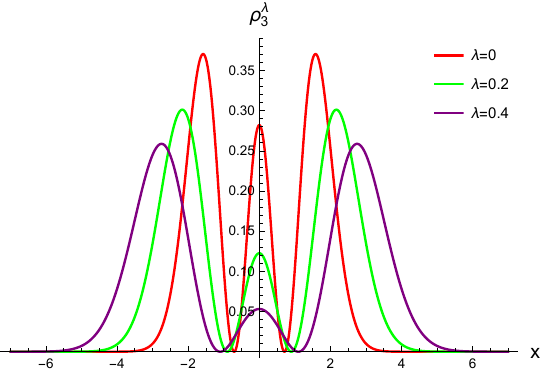}} &
\subfloat[]{\includegraphics[scale = 0.9]{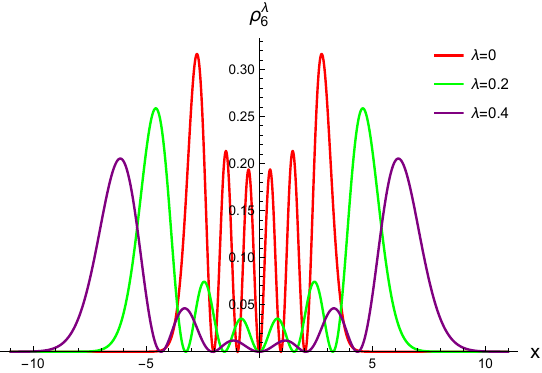}}
\end{tabular}
\caption{Density function in position space for several values of $\lambda$ and $n=0$(A), $n=1$(B), $n=3$(C) and $n=6$(D).}
\label{grid: density position}
\end{figure}

Hamiltonians like (\ref{eq: H class}), where the kinetic energy depends on position, admit multiple quantisation schemes. In general, different operator orderings lead to quantum Hamiltonians that are related by similarity transformations; as a result, their eigenfunctions differ by gauge transformations, while their spectra remain unchanged. In our case, the quantum one-dimensional Darboux III oscillator Hamiltonian was initially defined in the form \cite{ballesteros2011new} 
\begin{align} \label{eq: H}
    \mathcal{\hat{H}}(\hat{x},\hat{p})=\mathcal{T}(\hat{p})+\mathcal{U}(\hat{x})=\frac{1}{2  (1+\lambda \hat{x}^2)} \hat{p}^2+\frac{\omega^2 \hat{x}^2}{2 (1+\lambda \hat{x}^2)},
\end{align}
where the function depending on the coordinates is located at the left within the kinetic energy term. A detailed discussion on the ordering issues for this system can be found in \cite{ballesteros2011quantum} and references therein, where an equivalent Hamiltonian can be written
\begin{align}
    \hat{\mathcal{H}}_{\text{TPDM}}(\hat{x},\hat{p}) &= \frac{1}{2(1 + \lambda \hat{x}^2)} \hat{p}^2 
+ \frac{\omega^2 \hat{x}^2}{2(1 + \lambda \hat{x}^2)}
+ \frac{i \hbar \lambda}{(1 + \lambda \hat{x}^2)^2} (\hat{x} \cdot \hat{p})
+ \frac{\hbar^2 \lambda \left( 1 - 2 \lambda \x^2 \right)}{2(1 + \lambda \x^2)^3}.
\label{htpdm}
\end{align}
This new equivalent Hamiltonian has led to the analytical expressions for the Darboux III oscillator that will be used in subsequent Sections. In fact, analytical expressions for the eigenvalues $E_n^\lambda$ and for the wave function in position space $\Psi_n^\lambda(x)$ of (\ref{htpdm}) have been given in \cite{ballesteros2023shannon}. The eigenvalues are real and take the form 
\begin{align} \label{eq: Enlambda}
    E_{n}^\lambda&=- \hbar^2 \lambda \paren{n+\frac{1}{2}}^2+\hbar \paren{n+\frac{1}{2}} \sqrt{\hbar^2 \lambda^2 \paren{n+\frac{1}{2}}^2+\omega^2},
    \qquad
    n=0,1,2, \dots , \infty,
\end{align}
and the corresponding wave functions read 
\begin{align} \label{eq: funcion onda beta}
    \Psi_n^\lambda(x)={\left( \frac{\beta^2}{\pi} \right)^{1/4} \sqrt{\frac{1}{2^n n!}} \sqrt{\frac{1}{1+(n+\frac{1}{2}) \frac{\lambda}{\beta^2}}}} \sqrt{1+\lambda x^2} \mathrm{e}^{-\frac{\beta^2 x^2}{2}} {H_{n}\left( \beta x \right)}, \hspace{1cm} \beta=\sqrt{\frac{\Omega_n^\lambda}{\hbar}},
\end{align}
where the constant $\Omega_{n}^\lambda$ is an effective frequency given by
\begin{align} \label{eq: Omega def}
\Omega_{n}^\lambda=\sqrt{\omega^2-2 \lambda E_{n}^\lambda}.
\end{align}
Note that in this expression of $\On$, a larger value of $\omega$ counter-effects the influence of $\lambda$. In units with $\hbar=1$, for simplification purposes, the wave function can be written as
\begin{align}
    \Psi_n^\lambda(x)=\mathcal{N}_\lambda \sqrt{1+\lambda x^2}\mathrm{e}^{-\frac{\Omega_{n}^\lambda x^2}{2}} {H_{n}\left(\sqrt{\Omega_{n}^\lambda} x \right)},
\end{align}
where $\mathcal{N}_\lambda$ is the normalisation constant
\begin{equation} \label{eq: N normalizacion}
    \mathcal{N}_\lambda=\left( \frac{\Omega_n^\lambda}{\pi}  \right)^{1/4} \sqrt{\frac{1}{2^n n!}} \sqrt{\frac{1}{1+(n+\frac{1}{2}) \frac{\lambda}{\Omega_n^\lambda}}}. 
\end{equation}
\begin{figure}[H]
\begin{tabular}{cccc}
\subfloat[]{\includegraphics[scale = 0.9]{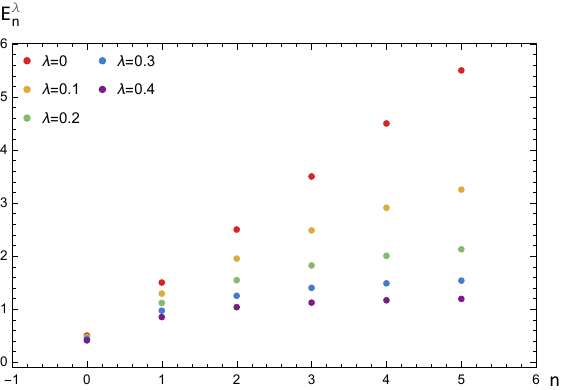}} &
\subfloat[]{\includegraphics[scale = 0.9]{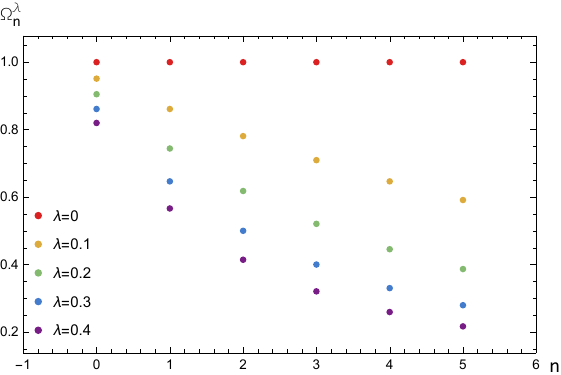}} 
\end{tabular}
\caption{Energy levels $E_n^\lambda$ (A) and effective frequency $\Omega_n^\lambda$ (B) vs $n$ for $\omega = 1$ and different values of $\lambda$ (indicated within the plot). The energy increases with the quantum number $n$ but decreases with $\lambda$. The frequency decreases with both. NUmerical data are given in Tables \ref{table: energy 1D}, \ref{table: Omega 1D}.}
\label{grid: Enlambda Omega}
\end{figure}
Accordingly, the density function in position space, $\rho_n^\lambda(x) = \abs{\Psi_n^\lambda(x)}^2$, is given by:
\begin{align} \label{eq: rho position}
\rho_n^\lambda(x)=\sqrt{\frac{\Omega_n^\lambda}{\pi}} \frac{1}{2^n n! \paren{1+(n+\frac{1}{2}) \frac{\lambda}{\Omega_n^\lambda}}  } \paren{1+\lambda x^2} \mathrm{e}^{-\Omega_{n}^\lambda x^2} {H^2_{n}\left(\sqrt{\Omega_{n}^\lambda} x \right)}.
\end{align}
As shown in Figure \ref{grid: density position}, increasing the values of $\lambda$,  as well as increasing values of $n$, will delocalise the probability density in position space.

\begin{figure}[H]
\begin{tabular}{cccc}
\subfloat[]{\includegraphics[scale = 0.9]{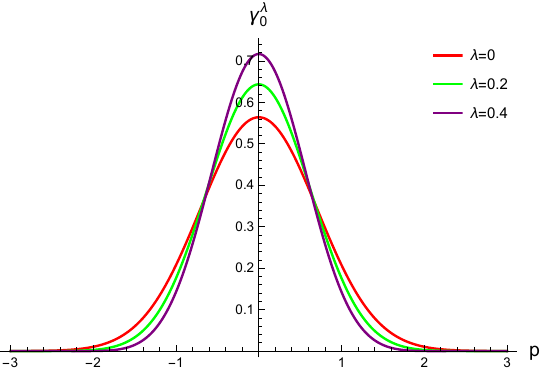}} &
\subfloat[]{\includegraphics[scale = 0.9]{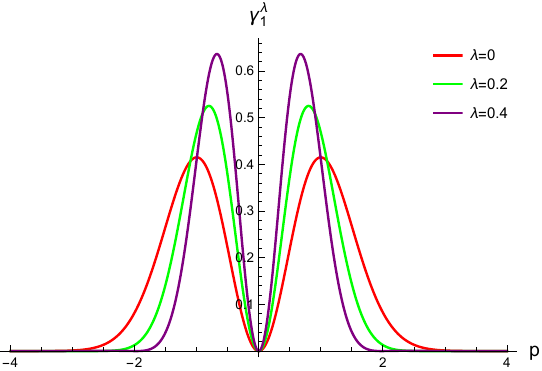}}  \\
\subfloat[]{\includegraphics[scale = 0.9]{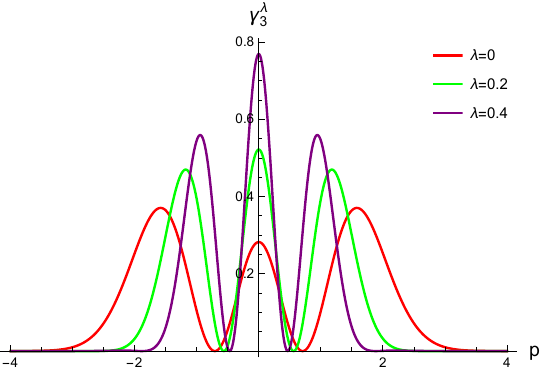}} &
\subfloat[]{\includegraphics[scale = 0.9]{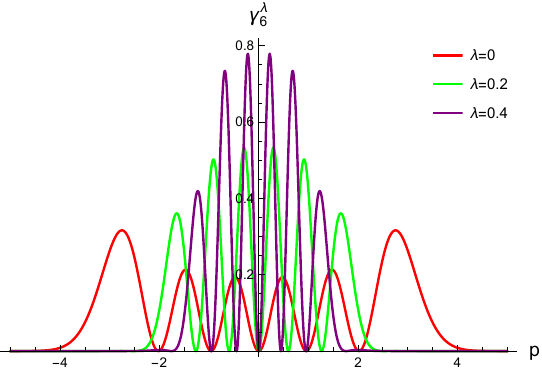}}
\end{tabular}
\caption{Density function in momentum space for several values of $\lambda$ and $n=0$(A), $n=1$(B), $n=3$(C) and $n=6$(D).}
\label{grid: density momentum}
\end{figure}

Wave functions in momentum space are given by the Fourier transform,
\begin{align} \label{eq: fourier trans 1D}
\Tilde{\Psi}_n^\lambda(p)=\mathcal{F} \llave{\Psi_{n}^\lambda(x)} =\frac{1}{\sqrt{2\pi}}\int_{\mathbb{R}} \expo{-ip x} \Psi_{n}^\lambda(x) \dif x.
\end{align}
Although the analytical wave function in position space is known, no closed-form expression can be obtained in momentum space from Eq.~\eqref{eq: fourier trans 1D}. Therefore, the density in momentum space  
\begin{align} \label{eq: density p space}
\gamma_n^\lambda(p)=\abs{\Tilde{\Psi}_n^\lambda(p)}^2=\frac{1}{2\pi} \abs{\int_{\mathbb{R}} \expo{-ip x} \Psi_{n}^\lambda(x) \dif x}^2\, ,    
\end{align}
was computed numerically. As expected, when the density delocalises in position space, it localises in momentum space with increasing $\lambda$, as shown in Figure \ref{grid: density momentum}. 

\section{Analytic entropic moments and Rényi and Tsallis entropies on position space}
\label{sec:position}

In this Section we study the entropic moments and the Rényi and Tsallis entropies for the position space wave-functions associated to the $1$-D Darboux III oscillator eigenstates. Remarkably enough, we are able to give analytical expressions in terms of the quantum number $n$ and of the nonlinearity parameter $\lambda$.

\subsection{Entropic moments}
Starting from the analytical expression (\ref{eq: rho position}) for the density function, we can write the entropic moments for a generic eigenstate as 
\begin{align}
     \mathcal{W}^{(\pe)} \left[\rho_n^\lambda \right] &=\int_{-\infty}^{+\infty} \paren{\rho_n^\lambda(x)}^{\pe} \dif x=   \mathcal{N}^{2\pe}_{\lambda}  \int_{-\infty}^{+\infty} \mathrm{e}^{-\Omega_{n}^\lambda \pe x^2} \paren{H_{n}\left(\sqrt{\Omega_{n}^\lambda} x \right)}^{2\pe} (1+\lambda x^2)^\pe \dif x,
  \label{paso: binomio theorem paso}
 \end{align}
 where $\mathcal{N}_{\lambda} $ is given by (\ref{eq: N normalizacion}). Since the results for the harmonic oscillator were derived in \cite{puertas2018exact} for $\pe \in \mathbb{N}$, we restrict our analysis to this case as well. The term $(1+\lambda x^2)^\pe$ in \eqref{paso: binomio theorem paso} can therefore be expanded using the binomial theorem. 
 \begin{align}    
      \mathcal{W}^{(\pe)} \left[\rho_n^\lambda \right] = \mathcal{N}^{2\pe}_{\lambda}  \sum_{k=0}^{\pe} \binom{\pe}{k} \lambda^k \int_{-\infty}^{+\infty} x^{2k} \mathrm{e}^{-\Omega_{n}^\lambda \pe x^2} \paren{H_{n}\left(\sqrt{\Omega_{n}^\lambda} x \right)}^{2\pe} \dif x.     
      \label{paso: integral de la suma}
\end{align}
Even powers of Hermite polynomials can be expressed as an infinite linear combination of Hermite polynomials (see \cite{puertas2018exact}), given by
\begin{align} \label{eq: trans hermite}
\paren{H_{n}\left(\sqrt{\Omega_{n}^\lambda} x \right)}^{2\pe}=A_{n,\pe}(\nu) \pe^{-\pe \nu}\sum_{j=0}^{\infty} \frac{c_{j,\pe}}{(-1)^j 2^{2j} j!} H_{2j}\left(\sqrt{\pe \ \Omega_{n}^\lambda} x \right),
\end{align}
where 
\begin{align} \label{eq: nu}
    \nu&=\frac{1}{2} (1-(-1)^n), \\
    \label{eq: A nu}
    A_{n,\pe}(\nu)&=2^{2\pe n} \paren{ \Gamma \paren{\frac{n-\nu}{2}+1}}^{2\pe}, \\ \notag
    c_{j,\pe}&=\left(\frac{1}{2} \right)_{\pe \nu} \binom{\frac{n+\nu-1}{2}}{\frac{n-\nu}{2}}^{2\pe} \sum_{j_{2\pe+1}=0}^j \sum_{j_1,\dots,j_{2\pe}}^{\frac{n-v}{2}} 
    \left(\pe \nu +\frac{1}{2} \right)_{j_1+\cdots+j_{2\pe}+j_{2\pe+1}} \frac{(-j)_{j_{2\pe+1}}}{(\frac{1}{2})_{j_{2\pe+1}} j_{2\pe+1}!} \\ 
    \label{eq: cj}
    &\frac{\left(\frac{n-\nu}{2} \right)_{j_1}}{\left(\nu +\frac{1}{2} \right)_{j_1}}
    \cdots
    \frac{\left(\frac{n-\nu}{2} \right)_{j_{2\pe}}}{\left(\nu +\frac{1}{2} \right)_{j_{2\pe}}}
    \frac{\left(\frac{1}{\pe} \right)^{j_1}}{j_1!} \cdots \frac{\left(\frac{1}{\pe} \right)^{j_{2\pe}}}{j_{2\pe}!},
\end{align}
where the Pochhammer's symbol
\begin{equation}
(z)_a=\frac{\Gamma(z+a)}{\Gamma(z)}
\, ,
\label{eq:pochhammer}
\end{equation}
appears in the last expression. 
 
Introducing \eqref{eq: trans hermite} into \eqref{paso: integral de la suma} and performing the change of variable $t=x \,\sqrt{\pe \ \Omega_{n}^\lambda} $ we get the following expression for the entropic moments 
\begin{align} \label{paso: entropic moment con integral}
     \mathcal{W}^{(\pe)} \corche{\rho_n^\lambda} &=\mathcal{N}^{2\pe}_{\lambda}  \sum_{k=0}^{\pe} \left( \binom{\pe}{k} \left(\frac{\lambda}{\pe  \Omega_{n}^\lambda}  \right)^k 
    \frac{A_{n,\pe}(\nu) \pe^{-\pe \nu}}{\sqrt{\pe \ \Omega_{n}^\lambda}} \sum_{j=0}^{\infty} \frac{c_{j,\pe}}{(-1)^j 2^{2j} j!}
     \int_{-\infty}^{\infty} t^{2k} \mathrm{e}^{-t^2} H_{2j}(t) \dif t \right) .    
\end{align}
Integrals of Hermite polynomials of the type appearing in this expression can be expressed in terms of the hypergeometric function (we use the conventions from \cite{abramowitz1968handbook}) 
\begin{align}
    \prescript{}{n}F_m\paren{a_1,a_2,\dots,a_n;b_1,b_2,\dots, b_m;z}=\sum_{j=0}^\infty \frac{\paren{a_1}_j \paren{a_2}_j \cdots \paren{a_n}_j}{\paren{b_1}_j \paren{b_2}_j \cdots \paren{b_m}_j} \frac{z^j}{j!},
\end{align}
where the Pochhammer's symbol \eqref{eq:pochhammer} has been used. In particular, the integrals from \eqref{paso: entropic moment con integral} only involve the particular case for $n=2, m=1$, for which the hipergeometric function reduces to 
\begin{align} \label{eq: Hypergeometric2F1}
    \prescript{}{2}F_1\paren{a_1,a_2;b;z}=\sum_{j=0}^\infty \frac{\paren{a_1}_j \paren{a_2}_j }{\paren{b}_j } \frac{z^j}{j!}.   
\end{align}
Explicitly (see \cite{prudnikov2018integrals}), integrals of Hermite polynomials of the type appearing in \eqref{paso: entropic moment con integral} read, for odd Hermite polynomials
\begin{align} 
\int_0^{\infty} x^{a-1} e^{-d x^2} H_{2 j+1}(c x) \dif x = (-1)^j \frac{2^{2 j}}{c^a} \left(\frac{2 - a}{2} \right)_j \Gamma\left(\frac{a + 1}{2} \right) 
\prescript{}{2}F_1\left( \frac{a}{2}, \frac{a + 1}{2} ; \frac{a}{2} - j ; \frac{c^2 - d}{c^2} \right),
\end{align}
whenever $\Re \, a > -1$ and $\Re \, d>0$, and for even Hermite polynomials
\begin{align} 
\int_0^{\infty} x^{a-1} e^{-d x^2} H_{2 j}(c x) \dif x = 
(-1)^j \frac{2^{2 j - 1}}{c^a} \left( \frac{1 - a}{2} \right)_j \Gamma\left( \frac{a}{2} \right) 
\prescript{}{2}F_1\left( \frac{a}{2}, \frac{a + 1}{2} ; \frac{a + 1}{2} - j ; \frac{c^2 - d}{c^2} \right),
\label{eq: conditions integral hermite 2j}
\end{align}
whenever $ \Re \, a > 0$ and $\Re \, d>0$. Using this last expression, since the integrals from \eqref{paso: entropic moment con integral} only involve even Hermite polynomials, and setting $a = 2 k + 1$, $d = 1$, $c=1$, we get
\begin{align} \notag
& \int_{-\infty}^{+\infty} t^{2k} e^{-t^2} H_{2j}(t) \dif t=2 \int_0^{\infty} t^{2k} e^{-t^2} H_{2j}(t) \dif t= \\
& =(-1)^j 2^{2 j}\paren{-k}_j \G{\frac{2k+1}{2}} \prescript{}{2}F_1\paren{\frac{2k+1}{2}, k+1 ; k+1-j ; 0} =(-1)^j 2^{2 j}\paren{-k}_j \G{\frac{2k+1}{2}} .
\end{align}
Finally, given that the Pochhammer symbol $(-k)_j$ is zero for any $j>k$, the infinite sum in the entropic moment (\ref{paso: entropic moment con integral}) reduces to a finite sum, and the analytic expression for the entropic moments of an arbitrary eigenstate of the $1$D Darboux III quantum nonlinear oscillator is given by  
\begin{align} \label{eq: entropic moment w}
     \mathcal{W}^{(\pe)} \corche{\rho_n^\lambda} &=\mathcal{N}^{2\pe}_{\lambda}  \frac{A_{n,\pe}(\nu) \pe^{-\pe \nu}}{\sqrt{\pe \ \Omega_{n}^\lambda}}   \sum_{k=0}^{\pe} \left \{ \binom{\pe}{k} \left(\frac{\lambda}{\pe  \Omega_{n}^\lambda}  \right)^k 
    \G{\frac{2k+1}{2}}\sum_{j=0}^{k} \frac{c_{j,\pe}}{j!}
    \paren{-k}_j  \right \} .    
\end{align}
It is interesting to analyse the particular cases of $\lambda = 0$ (harmonic oscillator) and $n=0$ (ground state) of this expression.
\begin{itemize}
    \item  Firstly, if in \eqref{eq: entropic moment w} we set $\lambda = 0$, the only term that contributes to the sum in $j$ is $j=0$. Moreover, $\Omega_n^\lambda$ goes to $\omega$, the Pochhamer symbol reduces to $(0)_0=1$, and $\Gamma \paren{\frac{1}{2}}=\sqrt{\pi}$. Finally, we obtain
\begin{equation}
	\mathcal{W}^{(\pe)} \corche{\rho_n^0} =\mathcal{N}^{2\pe} \sqrt{\frac{\pi}{\omega}}  \frac{A_{n,\pe}(\nu) \pe^{-\pe \nu}}{\sqrt{\pe \ \omega}}   c_{0,\alpha} . 
	\label{eq:W_lambda=0}
\end{equation}
Note that this expression for the entropic moment of a general eigenstate of the harmonic oscillator was obtained in \cite{puertas2018exact} for arbitrary dimension $D$ and gives that of the one-dimensional harmonic oscillator.
\item For the ground state $(n=0)$, the coefficients $c_{j,\pe}=\delta_{0,j}$, and thus the entropic moment \eqref{eq: entropic moment w} for the ground state of the one-dimensional Darboux III oscillator takes the form
\begin{align} \label{eq: W n=0}
     \mathcal{W}^{(\pe)} \left[\rho_0^\lambda \right] &=\paren{\frac{\Omega_0^\lambda}{\pi}}^{\frac{\pe-1}{2}} \frac{1}{\paren{1+\frac{1}{2} \frac{\lambda}{\Omega_0^\lambda}}^\pe  } \frac{1}{\sqrt{\pe \ \pi}}   \sum_{k=0}^{\pe}  \binom{\pe}{k} \left(\frac{\lambda}{\pe  \Omega_{0}^\lambda}  \right)^k 
    \G{\frac{2k+1}{2}} .
\end{align}
\item Furthermore, setting $\lambda = 0$ and $n=0$ we recover the well-known expression for the entropic moment of the ground state of the harmonic oscillator, which reads
\begin{align} \label{eq: W n=0, lambda=0}
     \mathcal{W}^{(\pe)} \left[\rho_0^0 \right] &=\paren{\frac{\omega}{\pi}}^{\frac{\pe-1}{2}}  \frac{1}{\sqrt{\pe}} .
\end{align}
\end{itemize}

One particularly interesting example of entropic moment is the case $\alpha=2$. This functional, known as the disequilibrium and denoted by $ \mathcal{D} \corche{\rho_n^\lambda} = \mathcal{W}^{(2)} \corche{\rho_n^\lambda} $, has been studied in connection to statistical complexity and thermodynamics \cite{pennini2017disequilibrium} and quantum entanglement of Rydberg multidimensional states \cite{dehesa2021momentum}. Setting $\alpha=2$ in Eq. \eqref{eq: entropic moment w},  we obtain analytic expressions for the disequilibrium of a general eigenstate of the one-dimensional Darboux III oscillator
\begin{align}
     \mathcal{D}  \corche{\rho_n^\lambda} &=\sqrt{\frac{\Omega_n^\lambda}{\pi}} \frac{1}{\paren{2^n n!}^2\paren{1+(n+\frac{1}{2}) \frac{\lambda}{\Omega_n^\lambda}}^2  } \frac{A_{n,2}(\nu) }{2^{2 \nu+\frac{1}{2}}} \paren{c_{0,2}+\frac{\lambda}{ \Omega_n^\lambda} \frac{c_{0,2}-c_{1,2}}{2}+\paren{\frac{\lambda}{ \Omega_n^\lambda}}^2 \frac{3 \paren{c_{0,2}-2 c_{1,2}+c_{2,2}}}{16}
     }.
\label{eq:Dnlambda}     
\end{align}
As relevant particular cases we mention:
\begin{itemize}
    \item The disequilibrium for a general eigenstate of the harmonic oscillator is obtain from \eqref{eq:W_lambda=0} and reads
\begin{align}
    \mathcal{D}  \corche{\rho_n^0} &=\sqrt{\frac{\omega}{\pi}} \frac{1}{\paren{2^n n!}^2\paren{1+(n+\frac{1}{2}) \frac{\lambda}{\omega}}^2  } \frac{A_{n,2}(\nu) }{2^{2 \nu+\frac{1}{2}}} c_{0,2} ,
\label{eq:Dn0}    
\end{align}
\item From \eqref{eq: W n=0}, the disequilibrium for the ground state of the one-dimensional Darboux III oscillator takes the form
\begin{equation}
\mathcal{D}  \corche{\rho_0^\lambda} =\frac{ \sqrt{\Omega_n^\lambda} \, \paren{3\lambda^2 + 8\lambda\Omega_n^\lambda + 16 \paren{\Omega_n^\lambda}^2} }{4 \sqrt{2\pi} \, \paren{\lambda + 2\Omega_n^\lambda}^2} .
\label{eq:D0lambda}
\end{equation}
\item Finally, setting $\alpha=2$ in \eqref{eq: W n=0, lambda=0} we recover the disequilibrium for the ground state of the harmonic oscillator, given by
\begin{equation}
\mathcal{D}  \corche{\rho_0^0} =\sqrt{\frac{\omega}{2\pi}} .
\label{eq:D00}
\end{equation}
\end{itemize}
Note that expressions \eqref{eq:Dn0} and \eqref{eq:D00} can be easily deduced from \cite{sanchez2010spreading}.

\subsection{Rényi and Tsallis entropies}
Once the entropic moments have been analytically computed for arbitrary values of the quantum number $n$ and the nonlinearity parameter $\lambda$, the Rényi and Tsallis entropies can be straightforwardly derived. For the Rényi entropy, after introducing \eqref{eq: entropic moment w} in \eqref{eq: renyi entropy} and performing some algebraic manipulations we get
\begin{align} \notag
     \mathcal{R}^{(\pe)} \left[\rho_n^\lambda \right] &=\frac{1}{2} \log{\frac{\pi}{\Omega_{n}^\lambda}}+\frac{1}{\pe-1} \log{\left(2^{\pe n}\pe^{\frac{1}{2}} \right)}+\frac{\pe}{\pe-1}
     \log{\left(1+\left(n+\frac{1}{2} \right) \frac{\lambda}{\Omega_n^\lambda} \right)} \\ \label{eq: entropia final renyi subs N}
     &+\frac{1}{1-\pe} \log{\paren{\frac{A_{n,\pe}(\nu)}{(n!)^\pe \sqrt{\pi} \pe^{\pe \nu}} \sum_{k=0}^\pe \eta_{\lambda}^{(k)}  }},
\end{align}
where $\eta_{\lambda}^{(k)}$ takes the form 
\begin{align} \label{eq: eta}
    \eta_{\lambda}^{(k)}&=       \binom{\pe}{k} \left(\frac{\lambda}{\pe  \Omega_{n}^\lambda}  \right)^k 
    \G{\frac{2k+1}{2}}\sum_{j=0}^{k} \frac{c_{j,\pe}}{j!}
    \paren{-k}_j .
\end{align}
Setting $\lambda = 0$ in the previous expression we get
\begin{align}
    \eta_{0}^{(k)}&=\sqrt{\pi}\, c_{0,\pe}  \, \delta_{k,0}\, ,
\end{align}
where $\delta_{k,0}$ is the Kronecker delta, since the only contributing term is the one with $k=j=0$. In this way we obtain the R\'enyi entropy for the one-dimensional harmonic oscillator (which was firstly presented in \cite{puertas2018exact}), which reads
\begin{align}
    \mathcal{R}^{(\pe)} \left[\rho_n^0 \right] &= \frac{1}{2} \log \frac{\pi}{\omega} + \frac{1}{\pe-1} \log \left( 2^{\pe n} \pe^{\frac{1}{2}} \right) + \frac{1}{1-\pe} \log \left( \frac{A_{n,\pe}(\nu)}{\pe^{\pe \nu} {(n!)^\pe}} c_{0,\pe} \right) .
\end{align}
We can also explicitly write the Rényi entropy for an arbitrary eigenstate of the one-dimensional Darboux III oscillator by setting $n=0$ in \eqref{eq: entropia final renyi subs N}, which gives 
\begin{equation}
\mathcal{R}^{(\pe)} \left[\rho_0^\lambda \right] =\frac{1}{2}\loga{\frac{\pi}{\Omega_0^\lambda}} +\frac{\pe}{\pe-1} \loga{1+\frac{1}{2} \frac{\lambda}{\Omega_0^\lambda}} +  \frac{1}{2\pe-2} \loga{\pe \pi}+   \frac{1}{1-\pe}\loga{\sum_{k=0}^{\pe}  \binom{\pe}{k} \left(\frac{\lambda}{\pe  \Omega_{0}^\lambda}  \right)^k 
    \G{\frac{2k+1}{2}} } .
\label{eq:R0lambda}
\end{equation}
Finally we recover the R\'enyi entropy for the ground state of the one-dimensional harmonic oscillator by setting both $\lambda=0$ and $n=0$, namely
\begin{equation}
\mathcal{R}^{(\pe)} \left[\rho_0^0 \right] =\frac{1}{2} \loga{\frac{\pi}{\omega}}+\frac{1}{2\pe-2} \log{\pe}.
\label{eq:R00}
\end{equation}

Similarly, analytical expressions for the Tsallis entropy can be derived from the entropic moments \eqref{eq: entropic moment w}. Explicitly, for a general eigenstate of the one-dimensional Darboux III oscillator we obtain 
\begin{align} \label{eq: Tsallis 1D darboux}
     \mathcal{T}^{(\pe)} \left[\rho_n^\lambda \right] &=\frac{1}{\pe-1} \llave{1-\paren{\frac{\Omega_n^\lambda}{\pi}}^{\pe/2} \frac{1}{\paren{2^n n!}^\pe \paren{1+(n+\frac{1}{2}) \frac{\lambda}{\Omega_n^\lambda}}^\pe  } \frac{A_{n,\pe}(\nu) \pe^{-\pe \nu}}{\sqrt{\pe \ \Omega_{n}^\lambda}}   \sum_{k=0}^{\pe} \eta^{(k)}_n} .
\end{align}
For $\lambda=0$, we get the Tsallis entropy for a general eigenstate of the one-dimensional harmonic oscillator, which reads
\begin{align}
     \mathcal{T}^{(\pe)} \left[\rho_n^0 \right] &=\frac{1}{\pe-1} \llave{1-\paren{\frac{\omega}{\pi}}^{\frac{\pe-1}{2}} \frac{1}{\paren{2^n n!}^\pe } \frac{A_{n,\pe}(\nu) }{\pe^{\pe \nu+\frac{1}{2}}} c_{0,\pe} } ,
\end{align}
while for $n=0$ we obtain the Tsallis entropy for the ground state of the one-dimensional Darboux III oscillator, which is given by
\begin{equation}
\mathcal{T}^{(\pe)} \left[\rho_0^\lambda \right] =\frac{1}{\pe-1} \paren{1-\paren{\frac{\Omega_0^\lambda}{\pi}}^{\frac{\pe-1}{2}} \frac{1}{\paren{1+\frac{1}{2} \frac{\lambda}{\Omega_0^\lambda}}^\pe  } \frac{1}{\sqrt{\pe \ \pi}}   \sum_{k=0}^{\pe}  \binom{\pe}{k} \left(\frac{\lambda}{\pe  \Omega_{0}^\lambda}  \right)^k 
    \G{\frac{2k+1}{2}}} .
\label{eq:T0lambda}
\end{equation}
Finally the Tsallis entropy for the ground state of the one-dimensional harmonic oscillator reads
\begin{equation}
\mathcal{T}^{(\pe)} \left[\rho_0^0 \right] =\frac{1}{\alpha - 1} \paren{ 1 - \paren{ \frac{\omega}{\pi} }^{\frac{\alpha - 1}{2}} \frac{1}{\sqrt{\alpha}} } .
\label{eq:T00}
\end{equation}

\subsection{Graphical representations in position space}

In the following we analyse the behaviour of the Rényi and Tsallis entropies with respect to the parameters $\lambda$, $n$ and $\alpha$.

\begin{figure}[h]
\begin{tabular}{cccc}
\subfloat[]{\includegraphics[scale = 0.9]{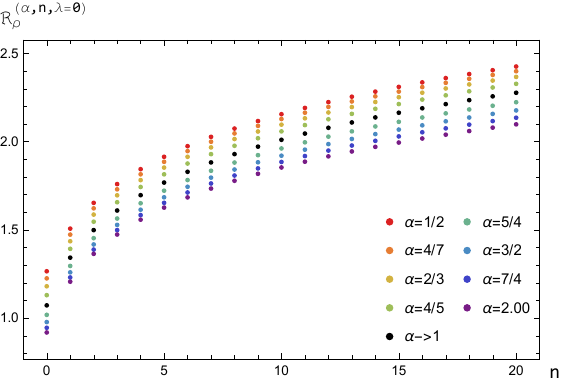}} &
\subfloat[]{\includegraphics[scale = 0.9]{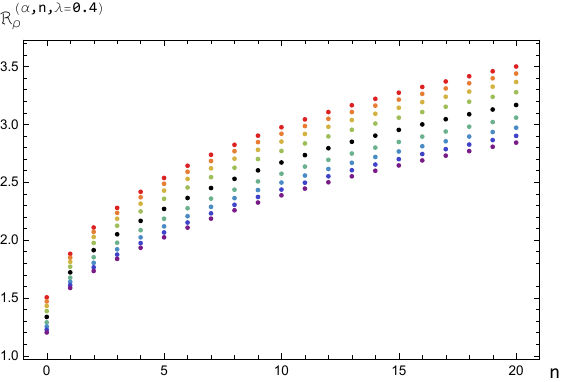}}  \\
\subfloat[]{\includegraphics[scale = 0.9]{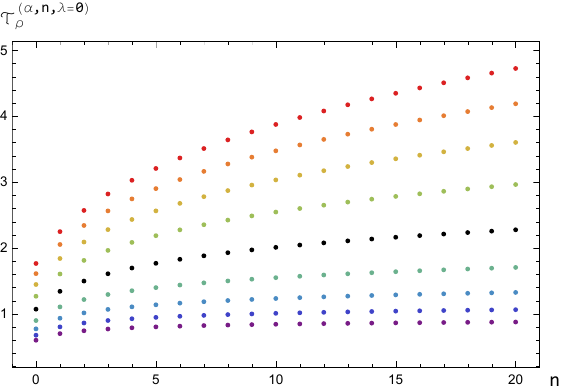}} &
\subfloat[]{\includegraphics[scale = 0.9]{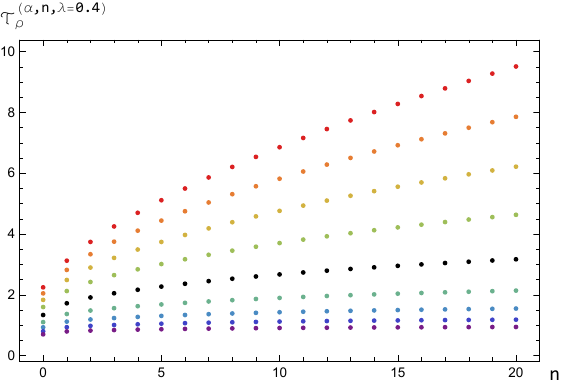}}
\end{tabular}
\caption{Effect of the parameter $\pe$: entropy in position space vs $n$ for several $\pe$ values given within panel (A). (A)\&(B) Rényi entropy $\mathcal{R}_\rho^{(\alpha,n,\lambda)}$, (C)\&(D) Tsallis entropy $\mathcal{T}_\rho^{(\alpha,n,\lambda)}$. (A)\&(C) Harmonic oscillator $(\lambda=0)$, (B)\&(D) Darboux III oscillator $(\lambda=0.4)$. Numerical data in Tables \ref{p2} (A), \ref{p3} (B), \ref{p4} (C), \ref{p5} (D).}
\label{grid: space p effect}
\end{figure}

Figure \ref{grid: space p effect} shows the dependence of the Rényi and Tsallis entropies on $n$ and the parameter $\pe$. To explore the behaviour both below and above the Shannon case ($\pe \to 1$), we selected a representative set of $\pe$ values. Let us comment on some of the main interesting features of Figure \ref{grid: space p effect}: 
\begin{itemize}
    \item As shown in Figure \ref{grid: density position}, and similarly to the harmonic oscillator case, the density spreads out as $n$ increases, leading to higher entropies, although the rate of increase diminishes with $n$.
    \item As expected from Eq.~\eqref{eq: shannon}, Rényi and Tsallis entropies approach the Shannon entropy when $\pe \to 1$, and their values vary according to relation (\ref{eq: p effect}). 
    \item For small values of $n$, the density is more localized in the harmonic oscillator case (panels A and C) than in the Darboux III oscillator with $\lambda = 0.4$ (panels B and D). This causes a sharper dependence on $\pe$ at low $n$.
    \item For higher values of $n$, the existence of $(n+1)$ maxima of the harmonic oscillator lead to a more uniform probability distribution than in the Darboux III oscillator, where the main peaks are located further from $x = 0$. As a result, the entropies of the Darboux III oscillator exhibit a stronger dependence on $\pe$ for large $n$.
    \item We also observe that, for the Darboux III oscillator with a sufficiently large $\lambda$, the entropies exceed those of the harmonic oscillator, especially for low values of $\pe$.
\end{itemize}

\clearpage
\noindent
\begin{multicols}{2}
  \begin{figure}[H] 
    \centering
    \subfloat[]{\includegraphics[scale=0.9]{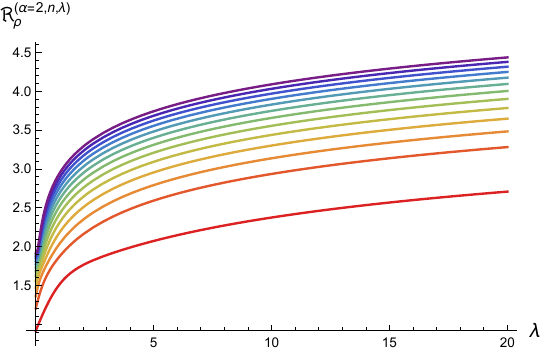}}\\[1ex]
    \subfloat[]{\includegraphics[scale=0.9]{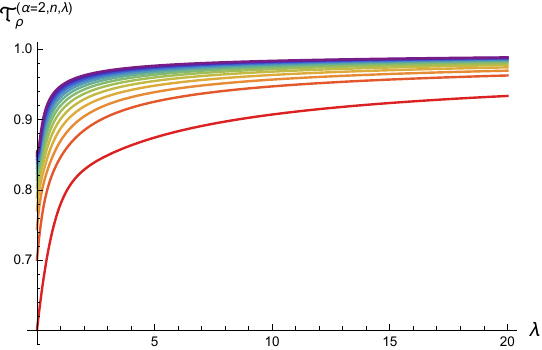}}\\[1ex]
    \subfloat[]{\includegraphics[scale=0.8]{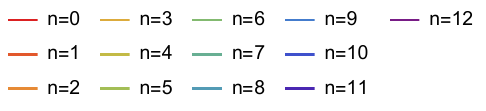}}
    \caption{Rényi $\mathcal{R}_\rho^{(\alpha,n,\lambda)}$ (\ref{eq: entropia final renyi subs N}) (A) and Tsallis $\mathcal{T}_\rho^{(\alpha,n,\lambda)}$ (\ref{eq: Tsallis 1D darboux}) (B) entropies in position space vs $\lambda$ (up to $\lambda=20$) for $\pe=2$, $\omega=1$, from $n=0$ to $n=9$ (C).}
    \label{grid: position lambda effect}
  \end{figure}

  \columnbreak

  \begin{figure}[H]
    \centering
    \subfloat[]{\includegraphics[scale=0.9]{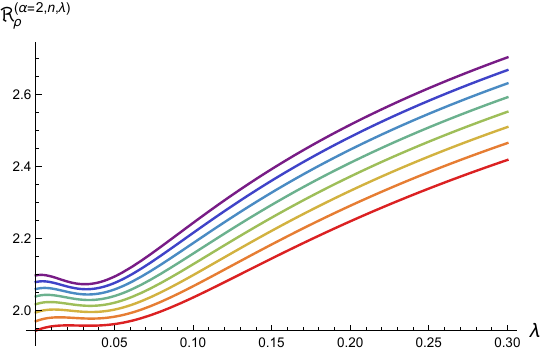}}\\[1ex]
    \subfloat[]{\includegraphics[scale=0.9]{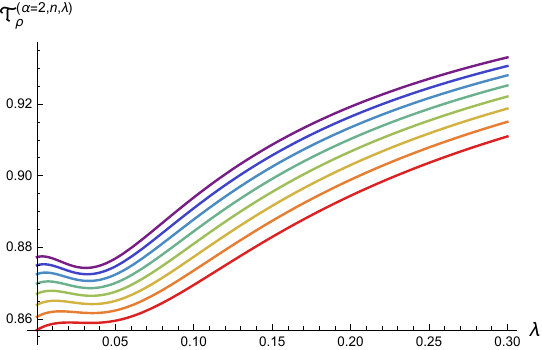}}\\[1ex]
    \subfloat[]{\includegraphics[scale=0.8]{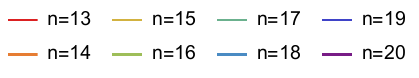}}
    \caption{Rényi $\mathcal{R}_\rho^{(\alpha,n,\lambda)}$ (\ref{eq: entropia final renyi subs N}) (A) and Tsallis $\mathcal{T}_\rho^{(\alpha,n,\lambda)}$ \eqref{eq: Tsallis 1D darboux} (B) entropies in position space vs $\lambda$ (up to $\lambda=0.30$) for $\pe=2$, $\omega=1$, from $n=13$ to $n=20$ (C).}
    \label{fig: continuous plot vs lambda position}
  \end{figure}
\end{multicols}

In the continuous plots of Figures \ref{grid: position lambda effect} and \ref{fig: continuous plot vs lambda position}, we observe the analytical effect of the nonlinearity parameter $\lambda$ on the Rényi and Tsallis entropies for a fixed value of $\pe$. Figure \ref{grid: position lambda effect} focuses on the Rényi and Tsallis entropies for the first excited states and reaching large values of $\lambda$:
\begin{itemize}
    \item As predicted, both entropies increase with $\lambda$. However, this is only true for all values of $\lambda$ if $n$ is sufficiently small.
    \item Entropies increase with $n$, but the curves become progressively closer for higher values of $n$, as anticipated from Figure \ref{grid: space p effect}.
    \item Tsallis entropy for large values of $\lambda$ (panel B of Figure \ref{grid: position lambda effect}) approaches 1, as the density becomes significantly delocalised. This is consistent with the discussion in the introduction regarding the limiting behaviour of the Tsallis entropy.
\end{itemize}

Figure \ref{fig: continuous plot vs lambda position} focuses on larger values of $n$, where a qualitatively different behaviour of the entropies emerges: for $n > 12$, they decrease before increasing again for certain small values of $\lambda$. This behaviour is explained by the interplay between the parameters $n$, $\pe$ and $\lambda$: 
\begin{itemize}
    \item The more excited the state, the more spread out the density becomes. However, the weight of the central part of the distribution decreases, as the most prominent maxima are located further from $x = 0$. 
    \item The parameter $\lambda$ also contributes to the spreading of the density, and further increases the relative dominance of the outermost peaks. Consequently, the central maxima may become negligible when compared to the other ones, depending on the value of $\pe$.
    \item For sufficiently large $n$, the central region holds a significant portion of the probability. A reduction in its relevance could outweigh the density's overall delocalization, leading to a decrease in entropy.
    \item Nevertheless, once the influence of the central region is outweighed, further increases in $\lambda$ primarily result in density spreading, and entropy increases again.
    \item Increasing the value of $\pe$ further filters the most probable part of the density. Therefore, the higher the value of $\pe$, the earlier this effect appears with increasing $n$.
\end{itemize}

\section{Rényi and Tsallis entropies on momentum space and entropic uncertainty relations}
\label{sec:momentum}

In the previous Section, we have obtained analytical expressions for the entropic moments and for the Rényi and Tsallis entropies for wave-functions in position space. In this Section, we first analyze the same quantities in momentum space and then study the entropic uncertainty relations associated with arbitrary eigenstates of the one-dimensional Darboux III nonlinear oscillator.

As already noted in Section \ref{sec:1D-DarbouxIII}, the Fourier transform \eqref{eq: fourier trans 1D} cannot be expressed in closed analytic form, and therefore an analytical treatment as in position space is not possible. For this reason, in the remainder of this Section we present a numerical analysis. 
\subsection{Rényi and Tsallis entropies in momentum space}

\begin{figure}[H]
\begin{tabular}{cccc}
\subfloat[]{\includegraphics[scale = 0.9]{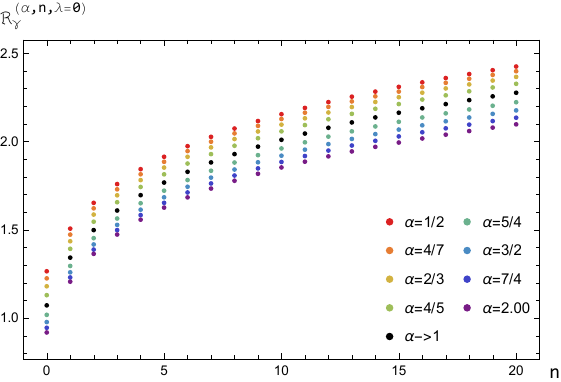}} &
\subfloat[]{\includegraphics[scale = 0.9]{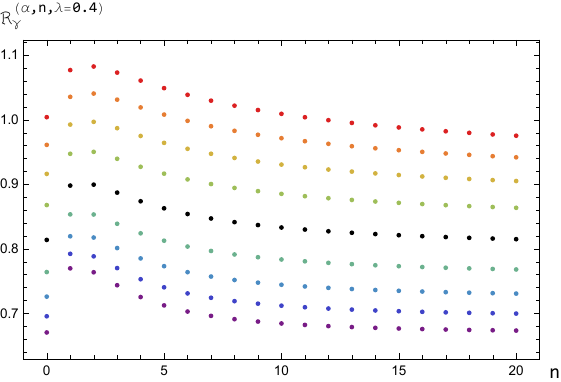}}  \\
\subfloat[]{\includegraphics[scale = 0.9]{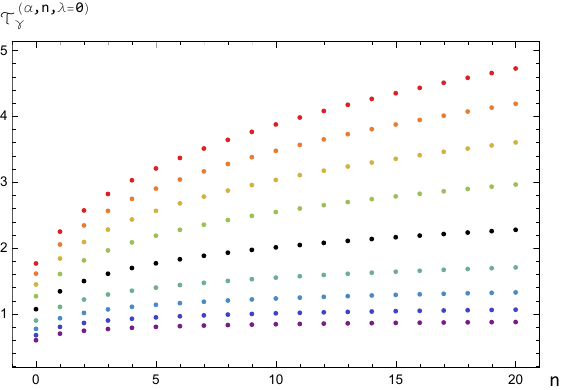}} &
\subfloat[]{\includegraphics[scale = 0.9]{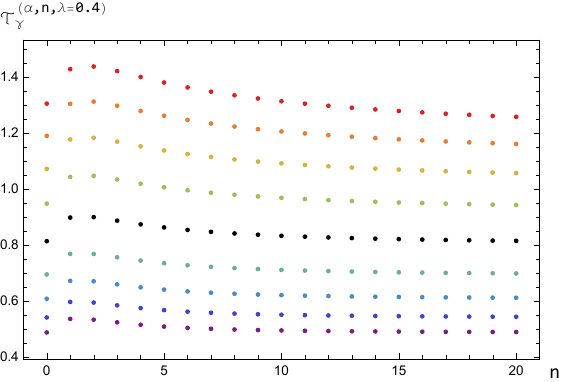}}
\end{tabular}
\caption{Effect of $\pe$: entropy in momentum space vs $n$ for several $\pe$ values described within panel (A). (A)\&(B) Rényi entropy $\mathcal{R}_\gamma^{(\alpha,n,\lambda)}$, (C)\&(D) Tsallis entropy $\mathcal{T}_\gamma^{(\alpha,n,\lambda)}$. (A)\&(C) Harmonic oscillator $(\lambda=0)$, (B)\&(D) Darboux III oscillator $(\lambda=0.4)$. Numerical data are given in Tables \ref{p9} (A), \ref{p10} (B), \ref{p11} (C) and \ref{p12} (D).}
\label{grid: momentum p effect}
\end{figure}

\begin{figure}[t]
\begin{tabular}{cccc}
\subfloat[]{\includegraphics[scale = 0.9]{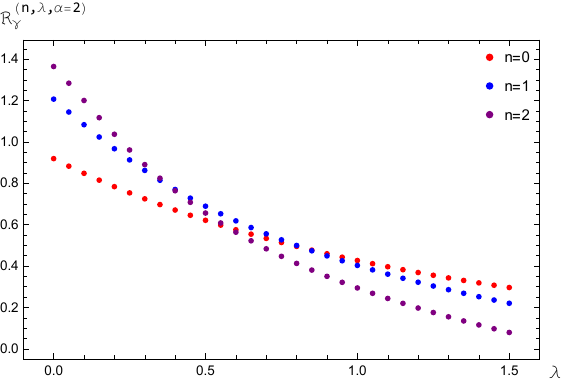}} &
\subfloat[]{\includegraphics[scale = 0.9]{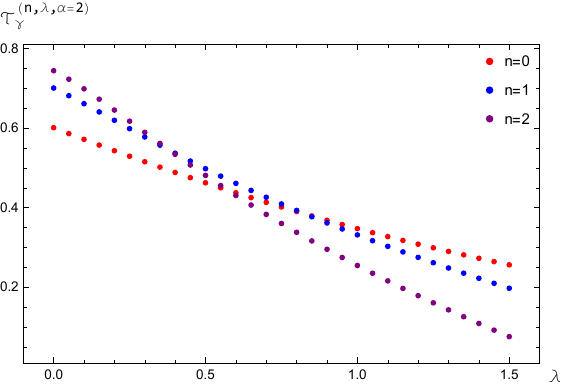}} 
\end{tabular}
\caption{Effect of $\lambda$: Rényi $\mathcal{R}_\gamma^{(\alpha,n,\lambda)}$ (A) and Tsallis $\mathcal{T}_\gamma^{(\alpha,n,\lambda)}$ (B) entropies in momentum space vs $\lambda$ for $n=0,1,2$, $\pe=2$ and $\omega=1$. Data in table \ref{table: momentum lambda effect}.}
  \label{grid: momentum lambda effect}
\end{figure}

In Figures \ref{grid: momentum p effect} and \ref{grid: momentum lambda effect} we present the numerical value of the Rényi and Tsallis entropies for the first eigenstates of the one-dimensional Darboux III nonlinear oscillator. 

In particular, Figure \ref{grid: momentum p effect} shows the dependence of the Rényi and Tsallis entropies in momentum space on $n$ and the parameter $\pe$: 
\begin{itemize}
    \item The dependence of the entropies on $\pe$ in momentum space closely resembles that in position space. Namely, the limiting value $\pe \to 1$ converges to the Shannon entropy and, as $\pe$ decreases, entropies increase.
    \item While both Rényi and Tsallis entropies increase monotonically with $n$ for the harmonic oscillator (panels A and C), this no longer holds for the Darboux III oscillator (panels B and D).
    \item In the Darboux III case, the entropies initially increase for small $n$, but then reach a maximum and begin to decrease.
    \item The $n$ value at which the entropy peaks depends on the parameter $\lambda$, and this behaviour is related to the entropy minimum observed in position space. We will discuss this further in Section~\ref{sec: large lambda}.
\end{itemize}

In Figure \ref{grid: momentum lambda effect}, we observe the effects of parameter $\lambda$ and quantum number $n$ for a fixed value $\pe=2$. In particular, we note that: 
\begin{itemize}
    \item Entropies decrease with $\lambda$ as the density becomes more localized. 
    \item This rate varies with the quantum number $n$ and, as a result, the entropy is initially higher for larger $n$ (as in the harmonic oscillator case), but ultimately becomes smaller. 
    \item This matches the decrease in entropy with $n$ observed in Figure~\ref{grid: momentum p effect}. 
\end{itemize}

\subsection{Entropy-based uncertainty principle}
For the Darboux III oscillator, the difference $\xi$ between the two sides of the uncertainty principle for the Rényi entropy \eqref{eq: zozor} and for the Tsallis entropy \eqref{eq: uncertainty Tsallis} reads, respectively 
\begin{align} \label{eq: xi R}
       \xi \corche{\mathcal{R}^{(\pe)} }&=\mathcal{R}^{(\pe)} \left[ \rho_n^\lambda(x)\right]+ \mathcal{R}^{(\beta)} \left[ \gamma_n^\lambda(p)\right] -
       \log{\left( \pi \pe^\frac{1}{2\pe-2} \beta^{\frac{1}{2\beta-2}}  \right)}, \hspace{1cm}
       \\ \label{eq: xi T}
       \xi \corche{\mathcal{T}^{(\pe)} }&=
        \paren{\frac{\pe}{\pi}}^{\frac{1}{4\pe}} \paren{\paren{1-\pe}\mathcal{T}^{(\pe)} \left[ \rho_n^\lambda\right]+1}^\frac{1}{2\pe} - \paren{\frac{\beta}{\pi}}^{\frac{1}{4\beta}} \paren{\paren{1-\beta}\mathcal{T}^{(\beta)} \left[ \gamma_n^\lambda \right]+1}^\frac{1}{2 \beta},     
\end{align}
where $\frac{1}{\pe}+\frac{1}{\beta} = 2$ in both expressions. In the Tsallis case, we also impose the constraint $\frac{1}{2} < \pe \leq 1$. In what follows, we refer to $\xi$ as the uncertainty function associated with a given entropy functional $X$, such that $\xi\corche{X} \geq 0$, and equality is reached only when the uncertainty relation is saturated.

\begin{figure}
\begin{tabular}{cccc}
\subfloat[]{\includegraphics[scale = 0.9]{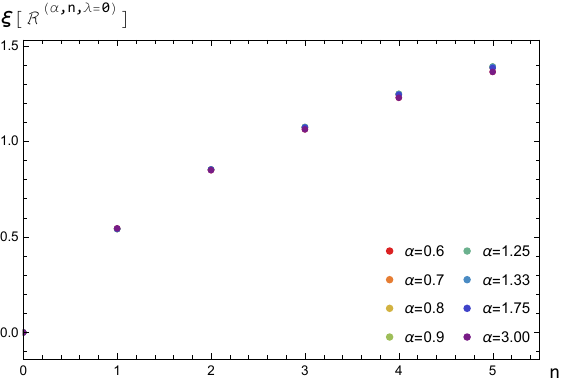}} &
\subfloat[]{\includegraphics[scale = 0.9]{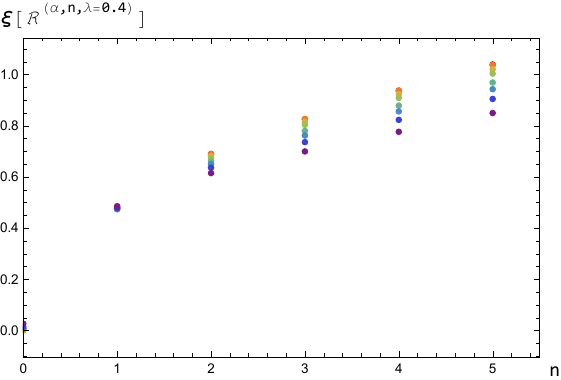}}  \\
\subfloat[]{\includegraphics[scale = 0.9]{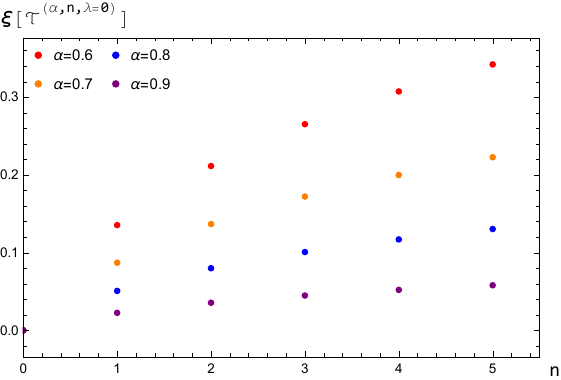}} &
\subfloat[]{\includegraphics[scale = 0.9]{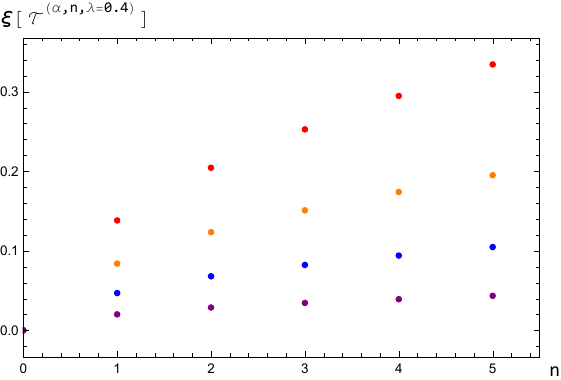}}
\end{tabular}
  \caption{Uncertainty function $\xi$   for: (A)\&(B) Rényi entropy $\mathcal{R}^{(\alpha,n,\lambda)}$, (C)\&(D) Tsallis entropy $\mathcal{T}^{(\alpha,n,\lambda)}$, where $\beta=1/\paren{2-1/\pe}$. (A)\&(C) harmonic oscillator $(\lambda=0)$, (B)\&(D) Darboux III Oscillator $(\lambda=0.4)$. Numerical data given in Tables \ref{p17} (A), \ref{p18} (B), \ref{p19} (C) and \ref{p20} (D).}
  \label{grid: uncertainty}
\end{figure}
Figure~\ref{grid: uncertainty} explores how the uncertainty function $\xi$ varies with $n$ and $\pe$:
\begin{itemize}
\item It is well known that the ground state of the harmonic oscillator (panels A and C) saturates the uncertainty relations \cite{puertas2018exact}, and therefore the uncertainty function vanishes.
\item The previous point is no longer true for the Darboux III oscillator (panels B and D). Owing to the limited resolution of the plots, this effect can be more clearly appreciated in the first rows of Tables \ref{p17} and \ref{p18}.
\item In all four panels, the uncertainty function increases with the quantum number $n$.
\item The observed decrease of $\xi$ with $\pe$ (and consequently, its increase with $\beta$) suggests that both Rényi and Tsallis entropies are more sensitive to variations of the entropic parameter in momentum space than in position space. This behaviour is consistent across both systems considered.
\item Comparing panels A and B, it is clear that $\xi$ is significantly more sensitive to changes in $\pe$ for the Darboux III oscillator. This difference, although it is also present in panels C and D, is less apparent in the case of the Tsallis entropy.

\end{itemize}

\begin{figure}[t]
\begin{tabular}{cccc}
\subfloat[]{\includegraphics[scale = 0.9]{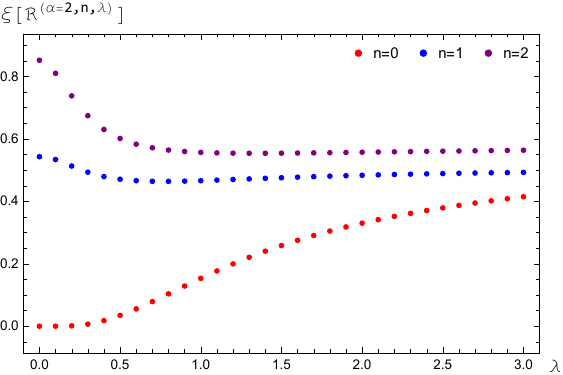}} &
\subfloat[]{\includegraphics[scale = 0.9]{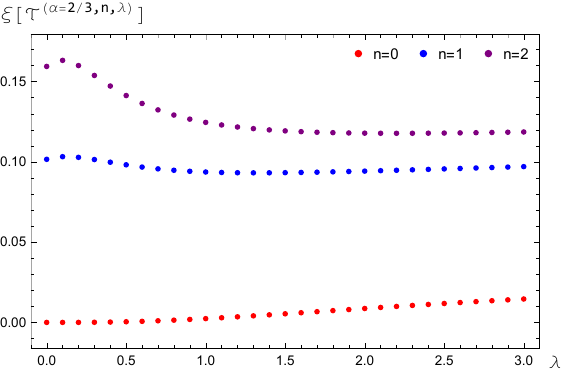}}
\end{tabular}
  \caption{$\xi$ function vs $\lambda$ for $\omega=1$. Rényi entropies for $\pe=2$ (A) and Tsallis entropies for $\pe=2/3$ (B). Data tables in \ref{table: uncertainty vs lambda A} and \ref{table: uncertainty vs lambda B}.}
  \label{grid: uncertainty vs lambda}
\end{figure}

Figure~\ref{grid: uncertainty vs lambda} explores the dependence of the uncertainty function on $n$ and $\lambda$, for fixed values of $\pe$. 
In panel A, the uncertainty increases with $\lambda$ for the ground state, and decreases with $\lambda$ for higher $n$. This behaviour was also observed in \cite{ballesteros2023shannon} for the Shannon entropy-based uncertainty relation. 
However, for certain values of $\pe$, for instance $\pe=2/3$ in panel B, the previously observed increase-then-decrease behaviour in entropy may also manifest in the uncertainties.


\section{Strong non-linear effects} \label{sec: large lambda}

In the previous Sections, we analyzed the entropic measures of the Darboux III oscillator. In particular, we described how increasing the nonlinearity  parameter $\lambda$ leads to density delocalization in position space and localization in momentum space. However, interesting effects emerge when  the quantum number $n$ is large enough, as observed in Figure~\ref{fig: continuous plot vs lambda position} and Figure~\ref{grid: momentum p effect}. In this Section, we examine more closely the interplay between $n$ and $\lambda$  in the highly excited state regime, i.e., for Rydberg-like states.

We recall that the radial entropy of highly energetic (i.e., Rydberg \cite{oks2021advances,gallagher1996production,gallagher1994rydberg}) states has been analytically derived for the harmonic oscillator \cite{dehesa2017entropic}. Even though the same results cannot be extended to the Darboux III oscillator, a different approach could still be employed to explore its behaviour in the Rydberg regime.

We begin by noting that the density function of an arbitrary Darboux III eigenstate can be separated into two terms: the first corresponds to the harmonic oscillator with frequency $\On$, and the second is a curvature-driven term that grows with $\lambda$ and contains $x^2$. Explicitly, we have
\begin{align} \notag
\rho_n^\lambda(x) =
\underbrace{\mathcal{N}_\lambda^2 e^{-\Omega_n^\lambda x^2} H_n^2\left(\sqrt{\Omega_n^\lambda} x\right)}_{{\text{harmonic-like term}}}
+
\underbrace{\lambda x^2 \mathcal{N}_\lambda^2 e^{-\Omega_n^\lambda x^2} H_n^2\left(\sqrt{\Omega_n^\lambda} x\right)}_{{\text{$\lambda-$induced term}}}=\rho_{n}^{(\lambda,0)}+\lambda \rho_{n}^{(\lambda,2)},
\end{align}
where we have defined
\begin{align}
\rho_{n}^{(\lambda,m)}=\mathcal{N}_\lambda^2 x^m \expo{-\Omega{n}^\lambda x^2} {H^2_{n}\paren{\sqrt{\Omega_{n}^\lambda} x}} .
\end{align}

In order to quantify the relative weigth of each term for given values of the parameters, we integrate each part of the density separately. For the first term we have
\begin{align} \label{eq: f}
f=\int_{-\infty}^{+\infty} \rho_{n}^{(\lambda,0)} \dx =
\frac{1}{1 + \paren{n + \frac{1}{2}} \frac{\lambda}{\On}},
\end{align}
and the second one is necessarily given by
\begin{align} \label{eq: 1-f}
1-f= \lambda \int_{-\infty}^{+\infty} \rho_{n}^{(\lambda,2)} \dx = \frac{1}{1 + \left( \paren{n + \frac{1}{2}} \frac{\lambda}{\On} \right)^{-1}}.
\end{align}

\begin{figure}[t]
\centering
\includegraphics[width=0.6\linewidth]{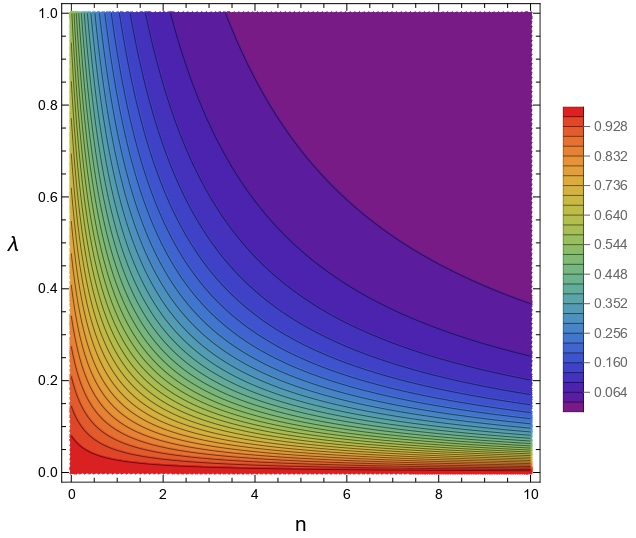}
\caption{Continuous contour plot of the function $f=\frac{1}{1 + \paren{n + \frac{1}{2}} \frac{\lambda}{\On}}$ vs $n$ and $\lambda$. The function approaches zero as either one or both variables increase and $\lambda \neq 0$.}
\label{fig: contorno}
\end{figure}

\begin{figure}[t]
\begin{tabular}{cccc}
\subfloat[]{\includegraphics[scale = 0.9]{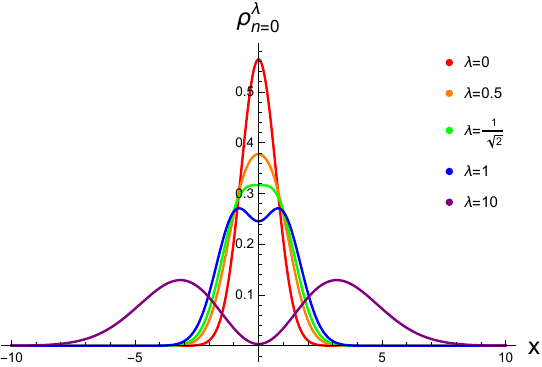}} &
\subfloat[]{\includegraphics[scale = 0.9]{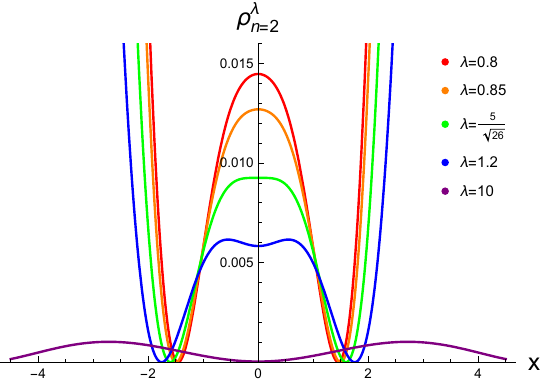}} \\
\subfloat[]
{\includegraphics[scale = 0.9]{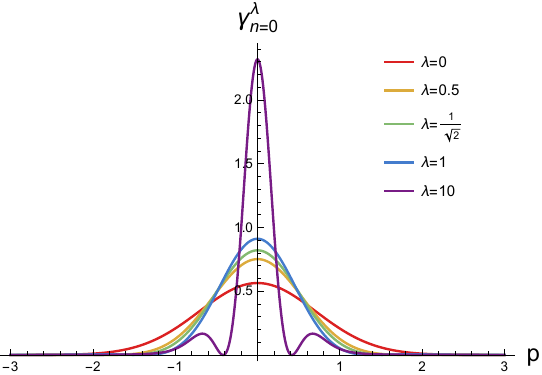}} & 
\subfloat[]{\includegraphics[scale = 0.9]{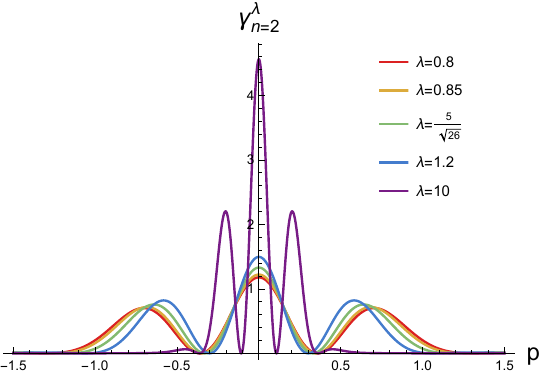}}
\end{tabular}
  \caption{Density in position space (A) \& (B) and momentum space (C) \& (D) for increasing different values of $\lambda$ (given within each panel) for $n=0$ (A) \& (C) and $n=2$ (B) \& (C) (zoom in (B) around $x=0$).}
  \label{grid: density limit lambda}
\end{figure}

Figure \ref{fig: contorno} shows that the harmonic-like term contribution becomes negligible as $f$ approaches zero, which occurs when the state becomes either more curved or more excited. In Figure \ref{grid: density limit lambda}, we explore how this behaviour manifests in the probability densities as $\lambda$ increases. This Figure can be interpreted by considering the one-dimensional Darboux III oscillator as a harmonic oscillator with a position-dependent mass. Initially, the densities expand and compress as expected (red and orange), but then there is a transition region (green) from which additional maxima emerge. This transition is related to the behaviour of the entropies decreasing before increasing again in position space (Figure\ref{grid: position lambda effect}) and the opposite effect observed in momentum space (Figures \ref{grid: momentum p effect} and \ref{grid: momentum lambda effect}). This effect was more pronounced the more excited the state is, and, as shown in Figure \ref{fig: contorno}, the influence of $\lambda$ is amplified by the quantum number $n$.


If $\lambda$ increases further (purple), the system can be seen as becoming very massive (and therefore slow) outside the vicinity of $x=0$. The probability of finding the particle is consequently higher where it spends most of its time (where it moves slower). In momentum space, the probability compresses, and this is consistent with an oscillator that becomes slower for which  the probability of smaller momenta increases. However, for sufficiently large values of $\lambda$, two additional maxima appear in momentum space. These less likely but higher momentum values are attributed to the particle crossing the origin with boosted speed due to the reduced mass.

The harmonic oscillator probability density function presents $n$ minima and $(n+1)$ maxima. In the Darboux III oscillator, depending on the parity of $n$ this behaviour is altered. While for odd values of $n$ there are no new maxima, for even values of $n$ two additional maxima appear for all $\lambda$ greater than some threshold (given by the green wavefunction in Figure \ref{grid: density limit lambda}). These maxima can be  computed analytically by solving the equation
\begin{align}
\frac{e^{\On x^2}}{H_n\paren{\sqrt{\On}x} \mathcal{N}_\lambda}\dv{\rho_n^\lambda(x)}{x} = 0 ,
\end{align}
where the normalisation constant and exponential factor are included by convenience and we are canceling out the zeroes associated to the Hermite polynomials (i.e., the minima). Manipulating the previous expression leads to
\begin{align} \label{eq: extremos densidad}
4n\sqrt{\On} \left(\lambda x^2 + 1\right) H_{n-1}\left(x \sqrt{\On}\right) - 2x \left(\lambda \left(x^2 \On - 1\right) + \On \right) H_n\left(x \sqrt{\On}\right) = 0.
\end{align}
Note that this is a polynomial of degree $(n+3)$ and therefore presents $(n+3)$ roots (counting real and complex roots and multiplicities). While Eq.~\eqref{eq: extremos densidad} cannot be solved analytically for arbitrary $n$, each individual case admits an exact analytic solution. 

Let us analyse the cases $n=0$ and $n=2$ in detail (see Figure \ref{grid: density limit lambda}): 

\begin{itemize}
    \item $n=0$: In this case the solutions of \eqref{eq: extremos densidad} are given by $x_0=0$ and $x_\pm=\pm \frac{\sqrt{\lambda -\Ona }}{\sqrt{\lambda \Ona } }$. Taking into consideration the definition of $\Omega_n^\lambda$ (from Eq. \eqref{eq: Omega def}) for $n=0$, these solutions are real only if $\lambda \geq \frac{\omega}{\sqrt{2}}$. Moreover, we can analyze the nature of these points by taking the second derivative and evaluating it at $x = 0$:
    \begin{align}
        \frac{1}{\mathcal{N}_\lambda}\dv[2]{\rho_0^\lambda}{x}&=\expo{-x^2 \Ona} \left( 2 \lambda + 4 x^2 \paren{\Ona}^2 \left( \lambda x^2 + 1 \right) - 2 \Ona \left( 5 \lambda x^2 + 1 \right) \right), \\ \left.\frac{1}{\mathcal{N}_\lambda}\dv[2]{\rho_0^\lambda}{x}\right|_{x=0}
        &=2 \paren{\lambda-\Ona}  .
    \end{align}
    The equation $\lambda = \Ona$ has the solution $\lambda=\frac{\omega}{\sqrt{2}}$. For this reason, the point $x=0$ is a maximum for $\lambda < \frac{\omega}{\sqrt{2}}$, an undulation point if $\lambda = \frac{\omega}{\sqrt{2}}$ and a minimum if $\lambda > \frac{\omega}{\sqrt{2}}$. Moreover, the extra critical points can also be studied in these exact three cases:   
    \begin{align}
       x_+ 
\begin{cases}
    \notin \mathbb{R}          & \text{if } \lambda < \frac{\omega}{\sqrt{2}} \\
  =0     & \text{if } \lambda=\frac{\omega}{\sqrt{2}} \\
  >0 & \text{if } \lambda > \frac{\omega}{\sqrt{2}}
\end{cases} 
,\hspace{0.8cm}
 x_-
\begin{cases}
    \notin \mathbb{R}          & \text{if } \lambda < \frac{\omega}{\sqrt{2}} \\
  =0     & \text{if } \lambda=\frac{\omega}{\sqrt{2}} \\
  <0 & \text{if } \lambda > \frac{\omega}{\sqrt{2}}
\end{cases} 
,\hspace{0.8cm}
\dv[2]{\rho_0^\lambda}{x} \bigg |_{x=x_\pm} \begin{cases}
  =0     & \text{if } \lambda=\frac{\omega}{\sqrt{2}} \\
  <0 & \text{if } \lambda > \frac{\omega}{\sqrt{2}}
\end{cases} .
    \end{align}
    As a conclusion, the central maximum turns into a minimum and two additional maxima appear when we surpass the threshold for $\lambda > \frac{\omega}{\sqrt{2}}$.
    \item $n=2$ : In this case, $$x_\pm=\pm \frac{1}{2} \sqrt{-\frac{\sqrt{41 \lambda ^2+12 \lambda  \Onc +4 \paren{\Onc}^2}}{\lambda  \Onc }-\frac{2}{\lambda }+\frac{7}{\Onc }}$$ are the new solutions. They are  real when $\lambda \geq \frac{5\omega}{\sqrt{26}}$ only. Let us perform the same analysis. The second derivative and its value in $x=0$ can be expressed as 
    \begin{align} \notag
       \frac{e^{x^2 \Onc}}{8\mathcal{N}_\lambda}\dv[2]{\rho_2^\lambda}{x}&=\lambda + \lambda x^2 \Onc \paren{8 x^6 \paren{\Onc}^3 - 60 x^4 \paren{\Onc}^2 + 98 x^2 \Onc - 29} \\ &+ \Onc \paren{8 x^6 \paren{\Onc}^3 - 44 x^4 \paren{\Onc}^2 + 46 x^2 \Onc - 5},
 \\ \left.\frac{1}{\mathcal{N}_\lambda}\dv[2]{\rho_2^\lambda}{x}\right|_{x=0}
        &=8 \paren{\lambda -5 \Onc}. 
    \end{align}
    The equation $\lambda=5 \Onc$ has the solution $\lambda=\frac{5\omega}{\sqrt{26}}$. 
    The point $x=0$ is a maximum for $\lambda <\frac{5\omega}{\sqrt{26}}$, an undulation point if $\lambda = \frac{5\omega}{\sqrt{26}}$ and a minimum if $\lambda > \frac{5\omega}{\sqrt{26}}$. The extra critical points are, depending on $\lambda$:   
    \begin{align}
        x_+ 
\begin{cases}
    \notin \mathbb{R}          & \text{if } \lambda < \frac{5\omega}{\sqrt{26}} \\
  =0     & \text{if } \lambda=\frac{5\omega}{\sqrt{26}} \\
  >0 & \text{if } \lambda > \frac{5\omega}{\sqrt{26}}
\end{cases} 
,\hspace{0.8cm} 
 x_- 
\begin{cases}
    \notin \mathbb{R}          & \text{if } \lambda < \frac{5\omega}{\sqrt{26}} \\
  =0     & \text{if } \lambda=\frac{5\omega}{\sqrt{26}} \\
  <0 & \text{if } \lambda > \frac{5\omega}{\sqrt{26}}
\end{cases}
, \hspace{0.8cm}
\dv[2]{\rho_n^\lambda}{x} \bigg |_{x=x_\pm} \begin{cases}
  =0     & \text{if } \lambda=\frac{5\omega}{\sqrt{26}} \\
  <0 & \text{if } \lambda > \frac{5\omega}{\sqrt{26}}
\end{cases}
.
    \end{align}
Therefore the behaviour is similar to the previous case: the central maximum turns into a minimum and two additional maxima appear when we surpass the threshold $\lambda > \frac{5 \omega}{\sqrt{26}}$. 
    
    
\end{itemize}

\begin{figure}[t]
\begin{tabular}{cccc}
\subfloat[]{\includegraphics[scale = 0.9]{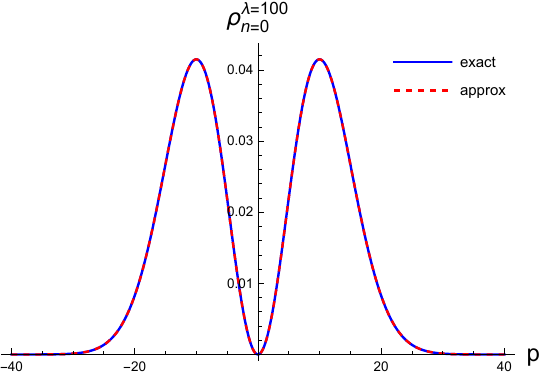}} &
\subfloat[]{\includegraphics[scale = 0.9]{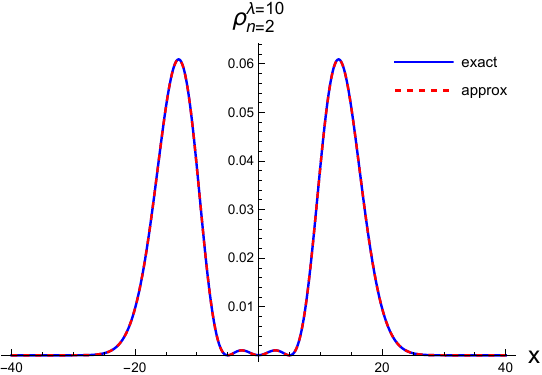}}
\end{tabular}
  \caption{Density representation in position space for $\omega=1$ for the exact expression (blue) in (\ref{eq: rho position}) and approximation $\lambda \rho_n^{\lambda,2}$ (red and dashed) for  $n=0$, $\lambda=100$ (A) and $n=2$, $\lambda=10$ (B). The values of $\lambda$ were not optimized, but rather chosen to qualitatively illustrate the approximation for an arbitrary large $\lambda$. The error of this approximation decreases for higher excited states, though the improvement becomes less significant for higher $n$. For this reason, the ground state requires a substantially high $\lambda$.}
  \label{fig: rho comparison}
\end{figure}

The value of $\lambda$ needed to reach the purple behaviour shown in Figure \ref{grid: density limit lambda} can also be determined. From Eq.~\eqref{eq: f}, one can choose an arbitrarily small $f$ to obtain this value of $\lambda$. In that case, the approximation $\rho_n^\lambda \approx \lambda \rho_n^{\lambda,2}$ becomes valid (Figure \ref{fig: rho comparison}). This approximation becomes increasingly accurate as $f$ approaches zero.

The wave function can be then approximated as $\varphi_n^\lambda$
\begin{align} \label{eq: approx lambda grande}
    \Psi_n^\lambda \approx \varphi_n^\lambda = \sqrt{\lambda \rho_n^{\lambda,2}}= \sqrt{\lambda} \mathcal{N}_\lambda \abs{x} \expo{-\On \frac{x^2}{2}} H_n \paren{\sqrt{\On} x},
\end{align}
for which analytical values for the Fourier transform can be found and read
\begin{align}
    \tilde{\varphi}_n^\lambda&=\frac{1}{\sqrt{2\pi}}\int_{-\infty}^{+\infty} \varphi_n^\lambda \expo{-i x p} \dx =\sqrt{\frac{\lambda}{2\pi} }\mathcal{N}_{\lambda} \int_{-\infty}^{+\infty} \abs{x} \expo{-\frac{\On x^2}{2}} H_{n}\paren{\sqrt{\On}x}\expo{-i x p} \dx\, , \\
    &=\sqrt{\frac{\lambda}{2\pi} } \frac{\mathcal{N}_{\lambda}}{\On} \int_{-\infty}^{+\infty} \abs{t} \expo{-\frac{t^2}{2}} H_{n}\paren{t}\expo{-i t P} \dt\, ,
\end{align}
with $P=\frac{p}{\sqrt{\On}}$. Separating this integral in the two intervals $(-\infty,0)$ and $(0,\infty)$, and changing from $t$ to $-t$ in the first one, we obtain
\begin{align}
    \tilde{\varphi}_n^\lambda&=\sqrt{\frac{\lambda}{2\pi} } \frac{\mathcal{N}_{\lambda}}{\On} \paren{\int_{-\infty}^{0} \paren{-t} \expo{-\frac{t^2}{2}} H_{n}\paren{t}\expo{-i t P} \dt+\int_{0}^{+\infty} t \expo{-\frac{t^2}{2}} H_{n}\paren{t}\expo{-i t P} \dt}, \\
    &=\sqrt{\frac{\lambda}{2\pi} } \frac{\mathcal{N}_{\lambda}}{\On} \paren{\int_{0}^{+\infty} \paren{-1}^n t \expo{-\frac{t^2}{2}} H_{n}\paren{t}\expo{i t P} \dt+\int_{0}^{+\infty} t \expo{-\frac{t^2}{2}} H_{n}\paren{t}\expo{-i t P} \dt}, \\ 
    &=\sqrt{\frac{\lambda}{2\pi} } \frac{\mathcal{N}_{\lambda}}{\On} \int_{0}^{\infty} t \expo{-\frac{t^2}{2}} H_{n}\paren{t} \paren{\expo{-i t P}+\paren{-1}^n \expo{i t P}} \dt, \\
    &=\sqrt{\frac{\lambda}{2\pi} } \frac{\mathcal{N}_{\lambda}}{\On} \int_{0}^{\infty} t \expo{-\frac{t^2}{2}} H_{n}\paren{t} \mathcal{P}\corche{\expo{-i t P}} \dt,  \\
    &=\sqrt{\frac{\lambda}{2\pi} } \frac{\mathcal{N}_{\lambda}}{\On} \mathcal{P}\corche{\int_{0}^{\infty} t \expo{-\frac{t^2}{2}} H_{n}\paren{t} \expo{-i t P}\dt },
\end{align}
with 
$\mathcal{P}\corche{\cdot}=2 \,\mathrm{Re} \corche{\cdot}$ if $n$ is even and $\mathcal{P}=-2 i \,\mathrm{Im} \corche{\cdot}$ otherwise. By substituting and making the change of variable  $t+i P=u$ we get
\begin{align}
    \tilde{\varphi}_n^\lambda &=\sqrt{\frac{\lambda}{2\pi} } \frac{\mathcal{N}_{\lambda}}{\On} \mathcal{P}\corche{\int_{0}^{\infty} t H_{n}\paren{t} \expo{-\frac{1}{2} \paren{t+i P}^2} \expo{- \frac{P^2}{2}} \dt }, \\
    &=\sqrt{\frac{\lambda}{2\pi} } \frac{\mathcal{N}_{\lambda}}{\On} \expo{- \frac{P^2}{2}} \mathcal{P}\corche{\int_{i P}^{\infty} (u-iP) H_{n}\paren{u-iP} \expo{-\frac{u^2}{2}}  \du }.
\end{align}
Using the translation property of the Hermite polynomials \cite{abramowitz1968handbook},
\begin{align}
    H_n(x+y)=\sum_{k=0}^n \binom{n}{k} 2^{n-k} H_k(x) y^{n-k},
\end{align}
the Fourier transform can be written as
\begin{align}
    \tilde{\varphi}_n^\lambda&=\sqrt{\frac{\lambda}{2\pi} } \frac{\mathcal{N}_{\lambda}}{\On} \expo{- \frac{P^2}{2}} \mathcal{P}\corche{\int_{iP}^{\infty} (u-iP) \paren{\sum_{k=0}^n \binom{n}{k} 2^{n-k} H_{k}\paren{-iP} u^{n-k}} \expo{-\frac{u^2}{2}}  \du }, \\
    &=\sqrt{\frac{\lambda}{2\pi} } \frac{\mathcal{N}_{\lambda}}{\On} \expo{- \frac{P^2}{2}} \mathcal{P}\corche{\sum_{k=0}^n \binom{n}{k} 2^{n-k} H_{k}\paren{-iP} \int_{iP}^{\infty} (u-iP)   u^{n-k} \expo{-\frac{u^2}{2}}  \du }.
\end{align}

The last integral takes the analytical form 
\begin{align}
    &\int_{iP}^{\infty} (u-iP) u^{n-k} \expo{-\frac{u^2}{2}} \, \mathrm{d}u 
    = 2^{\frac{n-k}{2}} \Gamma\left(\frac{n-k+2}{2}\right) \nonumber \\
    &\quad + 2^{\frac{n-k}{2}} ( P)^{n-k} \left(P^2\right)^{\frac{k-n}{2}} 
    \left( \Gamma\left(\frac{n-k+2}{2},-\frac{P^2}{2}\right)
    - \Gamma\left(\frac{n-k+2}{2}\right) \right) \nonumber \\
    &\quad - i P 2^{\frac{n-k-1}{2}} \Gamma\left(\frac{n-k+1}{2}\right)
    - i P 2^{\frac{n-k-1}{2}} (i P)^{1-(n-k)} \left(-P^2\right)^{\frac{k-n-1}{2}} \nonumber \\
    &\quad \times \left( \Gamma\left(\frac{n-k+1}{2},-\frac{P^2}{2}\right)
    - \Gamma\left(\frac{n-k+1}{2}\right) \right) =: g_{n,k}(P)\, ,
\end{align}
and  
\begin{align}
\Gamma \paren{s, x} = \int_x^\infty t^{s-1} e^{-t} \, dt \quad \text{for} \; s > 0 \; \text{and} \; x \geq 0 \, ,
\end{align}
is the incomplete gamma function \cite{abramowitz1968handbook}. The wave function in momentum space is then 
\begin{align} \label{eq: psi suma}
    \tilde{\varphi}_n^\lambda&=\sqrt{\frac{\lambda}{2\pi} } \frac{\mathcal{N}_{\lambda}}{\On} \expo{- \frac{P^2}{2}} \mathcal{P}\corche{\sum_{k=0}^n \binom{n}{k} H_{k}\paren{-iP} g_{n,k}(P) }.
\end{align}

\begin{figure}[t]
\begin{tabular}{cccc}
\subfloat[]{\includegraphics[scale = 0.9]{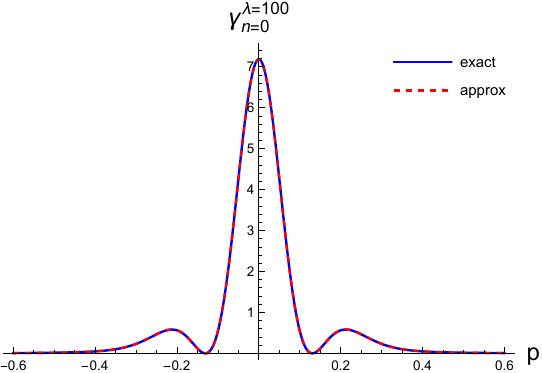}} &
\subfloat[]{\includegraphics[scale = 0.9]{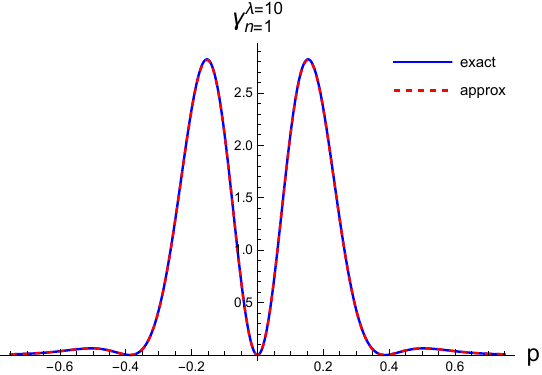}} \\
\subfloat[]
{\includegraphics[scale = 0.9]{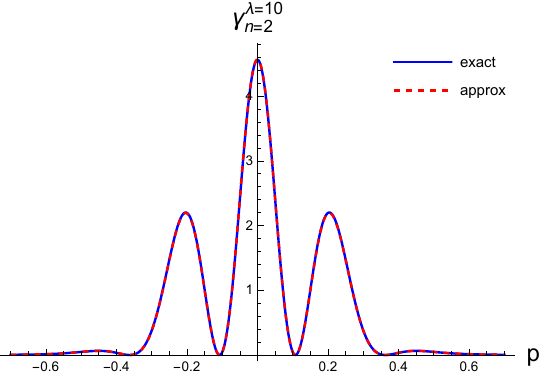}}&
\subfloat[]
{\includegraphics[scale = 0.9]{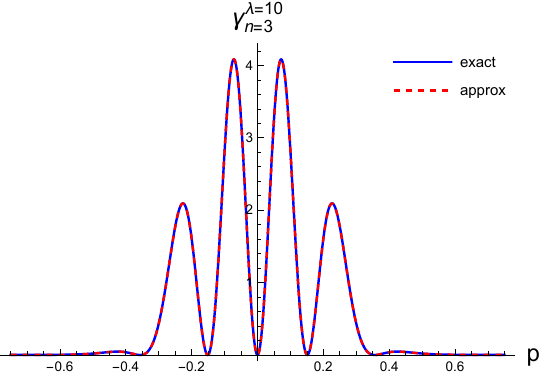}}
\end{tabular}
  \caption{Density representation in momentum space ($\omega=1$) for the numerical Fourier transform of the exact density in (\ref{eq: rho position}) (blue) and analytical Fourier transform in (\ref{eq: psi suma}) of the approximated wave function $\varphi_n^\lambda$ in (\ref{eq: approx lambda grande}) (red and dashed), with $n=0$ (\ref{eq: analytical Fourier transform n0}) (A), $n=1$  \eqref{eq: analytical Fourier transform n1} (B), $n=2$  \eqref{eq: analytical Fourier transform n2} (C) and $n=3$ \eqref{eq: analytical Fourier transform n3} (D). Nonlinearity parameter is $\lambda=100$ for $n=0$ (A) and $\lambda=10$ otherwise (B,C,D). These arbitrary large $\lambda$ were chosen following the same criteria as in Figure \ref{fig: rho comparison}.}
  \label{fig: gamma comparison}
\end{figure}   

Note that this function is real for even $n$ and imaginary for odd $n$, as expected from the parity of the integral \eqref{eq: approx lambda grande}. We can particularise this expression for $n=0,1,2,3$ obtaining the functions (which are plotted in Figure \ref{fig: gamma comparison}) 
\begin{align} \label{eq: analytical Fourier transform n0}
   \tilde{\Psi}_0^\lambda&= \frac{2 \sqrt{\frac{\lambda  \paren{\Ona}^{3/2}}{\lambda + 2 \Ona}} \left(\sqrt{\Ona} - \sqrt{2} p F\left(\frac{p}{\sqrt{2} \sqrt{\Ona}}\right)\right)}{\pi^{3/4} \paren{\Ona}^{3/2}},
\\ \label{eq: analytical Fourier transform n1}
      \tilde{\Psi}_1^\lambda&=-\frac{2 i \sqrt{\frac{\lambda  \paren{\Onb}^{3/2}}{3 \lambda +2 \Onb }}}{\pi ^{3/4} \paren{\Onb} ^2}  \left(2 \left(\Onb -p^2\right) F\left(\frac{p}{\sqrt{2 \Onb} }\right)+\sqrt{2} p \sqrt{\Onb }\right), \\ \label{eq: analytical Fourier transform n2}
      \tilde{\Psi}_2^\lambda&=\frac{\sqrt{\frac{\lambda  \paren{\Onc} ^{3/2}}{10 \lambda +4 \Onc }}}{\pi ^{3/4}} 
      \left(\frac{2 \sqrt{2} \left(2 p^3-5 p \Onc \right) F\left(\frac{p}{\sqrt{2 \Onc} }\right)}{\paren{\Onc} ^{5/2}}+\frac{6 \Onc -4 p^2}{\paren{\Onc}^2}\right), \\ \label{eq: analytical Fourier transform n3}
       \tilde{\Psi}_3^\lambda&=-\frac{2 i \sqrt{\frac{\lambda \paren{\Ond}^{3/2}}{21 \lambda + 6 \Ond}}}{\pi^{3/4} \paren{\Ond}^3} 
    \left(\sqrt{2} \left(2 p^4 - 9 p^2 \Ond + 3 \paren{\Ond}^2\right) F\left(\frac{p}{\sqrt{2 \Ond}}\right) + p \sqrt{\Ond} \left(7 \Ond - 2 p^2\right)\right),
\end{align}
where $F$ is the Dawson $F$ function defined \cite{abramowitz1968handbook} as
\begin{align}
    F(x) = e^{-x^2} \int_0^x e^{t^2} \, \dt.
\end{align}

In general, Eq. (\ref{eq: psi suma}) takes the form
\begin{equation}
\tilde{\varphi}_n^\lambda = \frac{\sqrt{\lambda \, \paren{\On}^{3/2}}}{\left( a_n \lambda + b_n \On \right) \, \pi^{3/4} \, \paren{\On}^{n + 1/2}} 
\paren{P_n(p, \On) + Q_n(p, \On) \, F\left( \frac{p}{\sqrt{2\On}} \right)},
\end{equation}
where $a_n$ and $b_n$ are polynomials in $n$, and $P_n$ and $Q_n$ are polynomials in $p$ and $n$, all of which arise from the real and imaginary parts taken according to the parity of $n$. These real or imaginary parts depend on products of complex functions that are either real or imaginary depending on the parity of $k$, which varies in each term of the summation.  
All in all, for the highly nonlinear regime and/or highly excited states, a closed analytical expression that provides a very good approximation (Figure \ref{fig: gamma comparison}) of the Fourier transform has been obtained for any $n$. This approximation can be made more accurate by making $f$ smaller in \eqref{eq: f}.

\section{Concluding remarks} \label{sec: conclusiones}


In this paper, we have studied the entropies of the one-dimensional Darboux III oscillator \eqref{eq: H}, which, as discussed in Section \ref{sec:1D-DarbouxIII}, can be seen as a nonlinear  oscillator endowed with a position-dependent mass such that in the limit $\lambda \to 0$ we recover the usual harmonic oscillator Hamiltonian. The main results we presented consist in the explicit analytical computations of the entropic moments and the Rényi and Tsallis entropies in position space and a complete numerical study of the same quantities in momentum space, together with an approximate analytical approach to the latter for large $\lambda$ and/or higher excited states. In this concluding Section, we comment on the most remarkable of these results and outline some future work. 

Firstly, the entropic moment $\mathcal{W}^{(\pe)}$ (\ref{eq: entropic moment w}), as well as the Rényi $\mathcal{R}^{(\pe)}$ (\ref{eq: entropia final renyi subs N}) and Tsallis $\mathcal{T}^{(\pe)}$ (\ref{eq: Tsallis 1D darboux}) entropies, have been analytically derived in position space. As expected,  the harmonic oscillator results previously obtained in \cite{puertas2018exact} are always recovered in the limit $\lambda \to 0$. We recall that in  \cite{ballesteros2023shannon} it was shown that for the lowest eigenstates and certain values of the nonlinearity  parameter $\lambda$, the Shannon entropy increases in position space with $n$ and $\lambda$. However, the parameter $\pe$ in the Rényi and Tsallis entropies enables the identification of new features of the Darboux III eigenstates, which are not evident from the density's analytical expression. In particular, we have shown that increasing $\lambda$ makes the central part of the density increasingly negligible. Also, we have shown that when the region close to zero of the Hermite polynomials is wide enough (that is, for a sufficiently large $n$), the entropies can decrease with $\lambda$ before increasing again, as seen in Figure \ref{fig: continuous plot vs lambda position}.



Conversely, since no closed-form expression is available for the corresponding Fourier transform, the analysis in momentum space was initially carried out numerically. Regarding the behaviour of the Shannon entropy in momentum space, in \cite{ballesteros2023shannon} it was shown that for the first eigenstates and certain values of the nonlinearity parameter $\lambda$ it increases with $n$ but decreases with $\lambda$. In this paper, we have shown that again the Rényi and Tsallis entropies generally decrease with $\lambda$. However, a similar non-monotonic behaviour emerges, as seen in Figure \ref{grid: momentum p effect}, where entropies first increase and then decrease with $n$. Moreover, in Figure \ref{grid: momentum lambda effect} we show that while entropies increase with $n$ for the harmonic oscillator, in the Darboux III oscillator they decrease with $n$ for sufficiently large $\lambda$. This is also related to the interplay between $\lambda$ and $n$: increasing the nonlinearity parameter $\lambda$ localises the momentum-space density more strongly, ultimately leading to lower entropies as $n$ increases.


As with the Shannon entropy \cite{ballesteros2023shannon}, uncertainty-based entropy measures given by (\ref{eq: zozor}) and (\ref{eq: uncertainty Tsallis}) increase with $\lambda$ for the ground state, but decrease for excited states (Figure \ref{grid: uncertainty vs lambda}A). However, by varying the value of $\pe$, the previously mentioned increase-then-decrease behaviour can also be observed in the uncertainty relations based on Rényi or Tsallis entropies (Figure \ref{grid: uncertainty vs lambda}B).

In addition, we have shown that, for large values of $\lambda$, the system behaves like a very slow and massive harmonic oscillator when far from $x = 0$, and like a very fast and light one near $x = 0$ (Figure \ref{grid: density limit lambda}). If $\lambda$ is large enough, thus leading to small $f$ in (\ref{eq: f}), the wave function can be approximated in the form (\ref{eq: approx lambda grande}) and its Fourier transform then admits a closed analytic form (\ref{eq: psi suma}). This approximation improves for higher energy states, and the value of $\lambda$ required for its validity decreases significantly as $n$ increases. These findings allow us to perform an  analytical study of this nonlinear oscillator in the highly excited state regime.


It is worth emphasizing that the Darboux III oscillator is an exactly solvable quantum model. It combines the advantage of having closed-form expressions for all energy levels and wave functions with the inclusion of a nonlinearity parameter $\lambda$ that admits a clear physical interpretation. In this work we have performed the analysis of $\lambda$-induced effects through both analytical and numerical tools, offering a richer landscape for studying quantum information measures in nonlinear quantum systems. This paves the way for ongoing and future work on this system, including local uncertainty measures \cite{BBGR2025dispersion}. Another worthwhile direction would be the study of the higher-dimensional Darboux III oscillator, where the parameter $\lambda$ encodes the curvature of the underlying conformally flat manifold. Finally, it would be also interesting to study the case when the nonlinearity parameter $\lambda$ is negative. All of these lines of research are work in progress and will be presented elsewhere.

\section*{Acknowledgements}

The authors acknowledge partial support from the grant PID2023-148373NB-I00 funded by MCIN/AEI /  10.13039/501100011033 / FEDER -- UE, and the Q-CAYLE Project funded by the Regional Government of Castilla y León (Junta de Castilla y León) and the Ministry of Science and Innovation MICIN through NextGenerationEU (PRTR C17.I1). I. Gutierrez-Sagredo thanks Universidad de La Laguna, where part of the work has been done, for the hospitality and support.


\bibliographystyle{abbrvurlmendeley}


\newgeometry{top=1cm, bottom=1.3cm, left=1.1cm, right=1.1cm}

\section{Appendix}





For the sake of completeness, in the following we include the Tables containing the numerical data that have been used to construct some the corresponding Figures presented in the body of the paper.

 \begin{table}[!ht]
    \centering
    \begin{tabular}{|c|c|c|c|c|c|c|}
    \hline
    \diagbox{\boldsymbol{$\lambda$}}{\boldsymbol{$n$}} & \textbf{0} & \textbf{1} & \textbf{2} & \textbf{3} & \textbf{4} & \textbf{5} \\ \hline
        \textbf{0} & 0.5 & 1.5 & 2.5 & 3.5 & 4.5 & 5.5 \\ \hline
        \textbf{0.1} & 0.47562 & 1.29178 & 1.95194 & 2.48318 & 2.90964 & 3.25199 \\ \hline
        \textbf{0.2} & 0.45249 & 1.11605 & 1.54508 & 1.82229 & 2.00413 & 2.12634 \\ \hline
        \textbf{0.3} & 0.43059 & 0.96988 & 1.25000 & 1.40000 & 1.48513 & 1.53658 \\ \hline
        \textbf{0.4} & 0.40990 & 0.84929 & 1.03553 & 1.12163 & 1.16607 & 1.19135 \\ \hline
    \end{tabular}
    \caption{Energy levels $E_n^\lambda$ for $\omega = 1$, $n=0,...5$ and different values of $\lambda$. (Data plotted in Figure \ref{grid: Enlambda Omega}A).}
    \label{table: energy 1D}
\end{table}

\begin{table}[!ht]
    \centering
    \begin{tabular}{|c|c|c|c|c|c|c|}
    \hline
        \diagbox{\boldsymbol{$\lambda$}}{\boldsymbol{$n$}} & \textbf{0} & \textbf{1} & \textbf{2} & \textbf{3} & \textbf{4} & \textbf{5} \\ \hline
        \textbf{0} & 1 & 1 & 1 & 1 & 1 & 1 \\ \hline
        \textbf{0.1} & 0.95125 & 0.86119 & 0.78078 & 0.70948 & 0.64659 & 0.59127 \\ \hline
        \textbf{0.2} & 0.90499 & 0.74403 & 0.61803 & 0.52066 & 0.44536 & 0.38661 \\ \hline
        \textbf{0.3} & 0.86119 & 0.64659 & 0.50000 & 0.40000 & 0.33003 & 0.27938 \\ \hline
        \textbf{0.4} & 0.81980 & 0.56619 & 0.41421 & 0.32047 & 0.25913 & 0.21661 \\ \hline
    \end{tabular}
    \caption{Effective frequency $\Omega_n^\lambda$ for $\omega = 1$, $n=0,...5$ and different values of $\lambda$. (Data plotted in Figure \ref{grid: Enlambda Omega} B).}
    \label{table: Omega 1D}
\end{table}

\begin{table}[!ht]
   \centering
   \begin{tabular}{|c|c|c|c|c|c|c|c|c|c|}
   \hline
  \diagbox{\boldsymbol{$n$}}{\boldsymbol{$\alpha$}}& \textbf{0.5} & \textbf{0.57} & \textbf{1} & \textbf{0.67} & \textbf{0.8} & \textbf{1.25} & \textbf{1.5} & \textbf{1.75} & \textbf{2} \\ \hline
       \textbf{0} & 1.266 & 1.225 & 1.181 & 1.130 & 1.072 & 1.019 & 0.978 & 0.945 & 0.919 \\ \hline
       \textbf{1} & 1.507 & 1.473 & 1.436 & 1.393 & 1.343 & 1.296 & 1.260 & 1.231 & 1.207 \\ \hline
       \textbf{2} & 1.653 & 1.622 & 1.587 & 1.546 & 1.499 & 1.453 & 1.417 & 1.388 & 1.364 \\ \hline
       \textbf{3} & 1.759 & 1.730 & 1.696 & 1.657 & 1.610 & 1.564 & 1.528 & 1.499 & 1.474 \\ \hline
       \textbf{4} & 1.844 & 1.815 & 1.782 & 1.743 & 1.697 & 1.651 & 1.614 & 1.584 & 1.558 \\ \hline
       \textbf{5} & 1.914 & 1.886 & 1.853 & 1.815 & 1.768 & 1.722 & 1.685 & 1.653 & 1.626 \\ \hline
       \textbf{6} & 1.974 & 1.946 & 1.914 & 1.876 & 1.829 & 1.782 & 1.744 & 1.712 & 1.684 \\ \hline
       \textbf{7} & 2.027 & 1.999 & 1.967 & 1.929 & 1.882 & 1.835 & 1.796 & 1.763 & 1.734 \\ \hline
       \textbf{8} & 2.074 & 2.047 & 2.015 & 1.976 & 1.929 & 1.881 & 1.842 & 1.808 & 1.778 \\ \hline
       \textbf{9} & 2.116 & 2.089 & 2.057 & 2.019 & 1.972 & 1.923 & 1.883 & 1.848 & 1.818 \\ \hline
       \textbf{10} & 2.155 & 2.128 & 2.096 & 2.058 & 2.010 & 1.961 & 1.920 & 1.885 & 1.854 \\ \hline
       \textbf{11} & 2.191 & 2.164 & 2.132 & 2.094 & 2.046 & 1.996 & 1.954 & 1.918 & 1.886 \\ \hline
       \textbf{12} & 2.224 & 2.197 & 2.165 & 2.127 & 2.078 & 2.028 & 1.986 & 1.949 & 1.916 \\ \hline
       \textbf{13} & 2.254 & 2.227 & 2.196 & 2.157 & 2.109 & 2.058 & 2.015 & 1.978 & 1.944 \\ \hline
       \textbf{14} & 2.283 & 2.256 & 2.225 & 2.186 & 2.137 & 2.086 & 2.042 & 2.004 & 1.970 \\ \hline
       \textbf{15} & 2.310 & 2.283 & 2.252 & 2.213 & 2.164 & 2.112 & 2.068 & 2.029 & 1.995 \\ \hline
       \textbf{16} & 2.336 & 2.309 & 2.277 & 2.238 & 2.189 & 2.137 & 2.092 & 2.053 & 2.018 \\ \hline
       \textbf{17} & 2.360 & 2.333 & 2.301 & 2.262 & 2.212 & 2.160 & 2.115 & 2.075 & 2.039 \\ \hline
       \textbf{18} & 2.383 & 2.356 & 2.324 & 2.285 & 2.235 & 2.182 & 2.137 & 2.096 & 2.060 \\ \hline
       \textbf{19} & 2.404 & 2.378 & 2.346 & 2.307 & 2.256 & 2.203 & 2.157 & 2.116 & 2.079 \\ \hline
       \textbf{20} & 2.425 & 2.399 & 2.367 & 2.327 & 2.277 & 2.223 & 2.177 & 2.135 & 2.098 \\ \hline
   \end{tabular}
   \caption{Rényi entropy for the harmonic oscillator in position space vs $n$ for $\omega=1$ and different values of $\pe$. (Data plotted in Figure \ref{grid: space p effect} A).}
   \label{p2}
\end{table}

\begin{table}[!ht]
   \centering
   \begin{tabular}{|c|c|c|c|c|c|c|c|c|c|}
   \hline
       \diagbox{\boldsymbol{$n$}}{\boldsymbol{$\alpha$}} & \textbf{0.5} & \textbf{0.57} & \textbf{1} & \textbf{0.67} & \textbf{0.8} & \textbf{1.25} & \textbf{1.5} & \textbf{1.75} & \textbf{2} \\ \hline
       \textbf{0} & 1.505 & 1.469 & 1.430 & 1.386 & 1.336 & 1.289 & 1.253 & 1.225 & 1.201 \\ \hline
       \textbf{1} & 1.881 & 1.848 & 1.811 & 1.770 & 1.721 & 1.674 & 1.639 & 1.610 & 1.586 \\ \hline
       \textbf{2} & 2.108 & 2.070 & 2.027 & 1.975 & 1.912 & 1.851 & 1.803 & 1.764 & 1.733 \\ \hline
       \textbf{3} & 2.278 & 2.235 & 2.184 & 2.124 & 2.050 & 1.977 & 1.920 & 1.874 & 1.837 \\ \hline
       \textbf{4} & 2.417 & 2.370 & 2.314 & 2.247 & 2.166 & 2.085 & 2.023 & 1.973 & 1.933 \\ \hline
       \textbf{5} & 2.537 & 2.487 & 2.428 & 2.356 & 2.269 & 2.184 & 2.118 & 2.066 & 2.023 \\ \hline
       \textbf{6} & 2.642 & 2.590 & 2.528 & 2.454 & 2.363 & 2.275 & 2.206 & 2.151 & 2.107 \\ \hline
       \textbf{7} & 2.737 & 2.683 & 2.620 & 2.543 & 2.450 & 2.358 & 2.287 & 2.231 & 2.185 \\ \hline
       \textbf{8} & 2.823 & 2.768 & 2.703 & 2.625 & 2.529 & 2.435 & 2.362 & 2.304 & 2.257 \\ \hline
       \textbf{9} & 2.902 & 2.846 & 2.780 & 2.700 & 2.602 & 2.507 & 2.432 & 2.373 & 2.324 \\ \hline
       \textbf{10} & 2.975 & 2.918 & 2.851 & 2.770 & 2.670 & 2.573 & 2.497 & 2.436 & 2.387 \\ \hline
       \textbf{11} & 3.043 & 2.985 & 2.917 & 2.835 & 2.734 & 2.635 & 2.558 & 2.496 & 2.445 \\ \hline
       \textbf{12} & 3.106 & 3.048 & 2.979 & 2.896 & 2.794 & 2.693 & 2.614 & 2.551 & 2.499 \\ \hline
       \textbf{13} & 3.165 & 3.107 & 3.037 & 2.953 & 2.849 & 2.748 & 2.668 & 2.603 & 2.551 \\ \hline
       \textbf{14} & 3.221 & 3.162 & 3.092 & 3.007 & 2.902 & 2.799 & 2.718 & 2.653 & 2.599 \\ \hline
       \textbf{15} & 3.273 & 3.214 & 3.143 & 3.058 & 2.952 & 2.848 & 2.766 & 2.699 & 2.645 \\ \hline
       \textbf{16} & 3.323 & 3.263 & 3.192 & 3.106 & 2.999 & 2.894 & 2.811 & 2.743 & 2.688 \\ \hline
       \textbf{17} & 3.371 & 3.310 & 3.239 & 3.152 & 3.044 & 2.938 & 2.854 & 2.786 & 2.729 \\ \hline
       \textbf{18} & 3.416 & 3.355 & 3.283 & 3.196 & 3.087 & 2.980 & 2.895 & 2.826 & 2.769 \\ \hline
       \textbf{19} & 3.459 & 3.398 & 3.326 & 3.238 & 3.128 & 3.020 & 2.934 & 2.864 & 2.806 \\ \hline
       \textbf{20} & 3.500 & 3.439 & 3.366 & 3.277 & 3.167 & 3.058 & 2.971 & 2.900 & 2.842 \\ \hline
   \end{tabular}
   \caption{Rényi entropy for the Darboux III oscillator $(\lambda=0.4)$ in position space vs $n$ for $\omega=1$ and different values of $\pe$. (Data plotted in Figure \ref{grid: space p effect} B).}
   \label{p3}
\end{table}

\begin{table}[!ht]
   \centering
   \begin{tabular}{|c|c|c|c|c|c|c|c|c|c|}
   \hline
       \diagbox{\boldsymbol{$n$}}{\boldsymbol{$\alpha$}} & \textbf{0.5} & \textbf{0.57} & \textbf{1} & \textbf{0.67} & \textbf{0.8} & \textbf{1.25} & \textbf{1.5} & \textbf{1.75} & \textbf{2} \\ \hline
       \textbf{0} & 1.766 & 1.611 & 1.447 & 1.268 & 1.072 & 0.899 & 0.773 & 0.677 & 0.601 \\ \hline
       \textbf{1} & 2.249 & 2.054 & 1.841 & 1.606 & 1.343 & 1.107 & 0.935 & 0.804 & 0.701 \\ \hline
       \textbf{2} & 2.571 & 2.342 & 2.091 & 1.812 & 1.499 & 1.218 & 1.015 & 0.863 & 0.744 \\ \hline
       \textbf{3} & 2.820 & 2.563 & 2.280 & 1.964 & 1.610 & 1.295 & 1.069 & 0.900 & 0.771 \\ \hline
       \textbf{4} & 3.028 & 2.746 & 2.433 & 2.086 & 1.697 & 1.353 & 1.108 & 0.927 & 0.789 \\ \hline
       \textbf{5} & 3.208 & 2.902 & 2.564 & 2.188 & 1.768 & 1.399 & 1.139 & 0.947 & 0.803 \\ \hline
       \textbf{6} & 3.367 & 3.040 & 2.678 & 2.276 & 1.829 & 1.438 & 1.164 & 0.964 & 0.814 \\ \hline
       \textbf{7} & 3.510 & 3.163 & 2.780 & 2.354 & 1.882 & 1.472 & 1.185 & 0.978 & 0.823 \\ \hline
       \textbf{8} & 3.641 & 3.276 & 2.872 & 2.424 & 1.929 & 1.501 & 1.204 & 0.990 & 0.831 \\ \hline
       \textbf{9} & 3.762 & 3.379 & 2.956 & 2.488 & 1.972 & 1.527 & 1.220 & 1.000 & 0.838 \\ \hline
       \textbf{10} & 3.875 & 3.475 & 3.034 & 2.546 & 2.010 & 1.550 & 1.234 & 1.009 & 0.843 \\ \hline
       \textbf{11} & 3.981 & 3.565 & 3.106 & 2.600 & 2.046 & 1.571 & 1.247 & 1.017 & 0.848 \\ \hline
       \textbf{12} & 4.080 & 3.649 & 3.174 & 2.650 & 2.078 & 1.591 & 1.259 & 1.024 & 0.853 \\ \hline
       \textbf{13} & 4.174 & 3.728 & 3.237 & 2.697 & 2.109 & 1.609 & 1.270 & 1.031 & 0.857 \\ \hline
       \textbf{14} & 4.264 & 3.803 & 3.297 & 2.742 & 2.137 & 1.625 & 1.280 & 1.037 & 0.861 \\ \hline
       \textbf{15} & 4.348 & 3.875 & 3.354 & 2.783 & 2.164 & 1.641 & 1.289 & 1.042 & 0.864 \\ \hline
       \textbf{16} & 4.430 & 3.943 & 3.408 & 2.823 & 2.189 & 1.655 & 1.297 & 1.047 & 0.867 \\ \hline
       \textbf{17} & 4.508 & 4.008 & 3.460 & 2.861 & 2.212 & 1.669 & 1.305 & 1.052 & 0.870 \\ \hline
       \textbf{18} & 4.583 & 4.071 & 3.510 & 2.897 & 2.235 & 1.682 & 1.313 & 1.057 & 0.873 \\ \hline
       \textbf{19} & 4.655 & 4.131 & 3.557 & 2.931 & 2.256 & 1.694 & 1.320 & 1.061 & 0.875 \\ \hline
       \textbf{20} & 4.725 & 4.189 & 3.603 & 2.964 & 2.277 & 1.705 & 1.326 & 1.065 & 0.877 \\ \hline
   \end{tabular}
   \caption{Tsallis entropy for the harmonic oscillator  in position space vs $n$ for $\omega=1$ and different values of $\pe$. (Data plotted in Figure \ref{grid: space p effect} C).}
   \label{p4}
\end{table}

\begin{table}[!ht]
   \centering
   \begin{tabular}{|c|c|c|c|c|c|c|c|c|c|}
   \hline
       \diagbox{\boldsymbol{$n$}}{\boldsymbol{$\alpha$}} & \textbf{0.5} & \textbf{0.57} & \textbf{1} & \textbf{0.67} & \textbf{0.8} & \textbf{1.25} & \textbf{1.5} & \textbf{1.75} & \textbf{2} \\ \hline
       \textbf{0} & 2.244 & 2.047 & 1.833 & 1.597 & 1.336 & 1.102 & 0.931 & 0.801 & 0.699 \\ \hline
       \textbf{1} & 3.122 & 2.818 & 2.487 & 2.123 & 1.721 & 1.368 & 1.119 & 0.935 & 0.795 \\ \hline
       \textbf{2} & 3.739 & 3.334 & 2.895 & 2.422 & 1.912 & 1.482 & 1.188 & 0.978 & 0.823 \\ \hline
       \textbf{3} & 4.248 & 3.748 & 3.214 & 2.646 & 2.050 & 1.560 & 1.234 & 1.006 & 0.841 \\ \hline
       \textbf{4} & 4.696 & 4.109 & 3.488 & 2.838 & 2.166 & 1.625 & 1.273 & 1.030 & 0.855 \\ \hline
       \textbf{5} & 5.109 & 4.440 & 3.738 & 3.010 & 2.269 & 1.683 & 1.306 & 1.050 & 0.868 \\ \hline
       \textbf{6} & 5.495 & 4.747 & 3.969 & 3.169 & 2.363 & 1.735 & 1.336 & 1.068 & 0.878 \\ \hline
       \textbf{7} & 5.859 & 5.036 & 4.184 & 3.315 & 2.450 & 1.782 & 1.363 & 1.083 & 0.888 \\ \hline
       \textbf{8} & 6.205 & 5.309 & 4.387 & 3.452 & 2.529 & 1.824 & 1.386 & 1.097 & 0.895 \\ \hline
       \textbf{9} & 6.536 & 5.569 & 4.578 & 3.580 & 2.602 & 1.862 & 1.407 & 1.108 & 0.902 \\ \hline
       \textbf{10} & 6.852 & 5.817 & 4.760 & 3.701 & 2.670 & 1.898 & 1.426 & 1.119 & 0.908 \\ \hline
       \textbf{11} & 7.157 & 6.054 & 4.933 & 3.815 & 2.734 & 1.930 & 1.443 & 1.128 & 0.913 \\ \hline
       \textbf{12} & 7.451 & 6.282 & 5.098 & 3.923 & 2.794 & 1.960 & 1.459 & 1.137 & 0.918 \\ \hline
       \textbf{13} & 7.735 & 6.501 & 5.256 & 4.025 & 2.849 & 1.988 & 1.473 & 1.144 & 0.922 \\ \hline
       \textbf{14} & 8.010 & 6.713 & 5.408 & 4.123 & 2.902 & 2.013 & 1.486 & 1.151 & 0.926 \\ \hline
       \textbf{15} & 8.276 & 6.918 & 5.554 & 4.217 & 2.952 & 2.037 & 1.498 & 1.157 & 0.929 \\ \hline
       \textbf{16} & 8.536 & 7.116 & 5.695 & 4.306 & 2.999 & 2.060 & 1.509 & 1.163 & 0.932 \\ \hline
       \textbf{17} & 8.788 & 7.308 & 5.831 & 4.392 & 3.044 & 2.081 & 1.520 & 1.168 & 0.935 \\ \hline
       \textbf{18} & 9.034 & 7.495 & 5.962 & 4.474 & 3.087 & 2.101 & 1.530 & 1.173 & 0.937 \\ \hline
       \textbf{19} & 9.274 & 7.676 & 6.090 & 4.554 & 3.128 & 2.120 & 1.539 & 1.178 & 0.940 \\ \hline
       \textbf{20} & 9.508 & 7.853 & 6.213 & 4.630 & 3.167 & 2.138 & 1.547 & 1.182 & 0.942 \\ \hline
   \end{tabular}
   \caption{Tsallis entropy for the Darboux III oscillator $(\lambda=0.4)$  in position space vs $n$ for $\omega=1$ and different values of $\pe$. (Data plotted in Figure \ref{grid: space p effect} D).}
   \label{p5}
\end{table}

\begin{table}[!ht]
   \centering
   \begin{tabular}{|c|c|c|c|c|c|c|c|c|c|}
   \hline
       \diagbox{\boldsymbol{$n$}}{\boldsymbol{$\alpha$}} & \textbf{0.5} & \textbf{0.57} & \textbf{1} & \textbf{0.67} & \textbf{0.8} & \textbf{1.25} & \textbf{1.5} & \textbf{1.75} & \textbf{2} \\ \hline
       \textbf{0} & 1.266 & 1.225 & 1.181 & 1.130 & 1.072 & 1.019 & 0.978 & 0.945 & 0.919 \\ \hline
       \textbf{1} & 1.507 & 1.473 & 1.436 & 1.393 & 1.343 & 1.296 & 1.260 & 1.231 & 1.207 \\ \hline
       \textbf{2} & 1.653 & 1.622 & 1.587 & 1.546 & 1.499 & 1.453 & 1.417 & 1.388 & 1.364 \\ \hline
       \textbf{3} & 1.759 & 1.730 & 1.696 & 1.657 & 1.610 & 1.564 & 1.528 & 1.499 & 1.474 \\ \hline
       \textbf{4} & 1.844 & 1.815 & 1.782 & 1.743 & 1.697 & 1.651 & 1.614 & 1.584 & 1.558 \\ \hline
       \textbf{5} & 1.914 & 1.886 & 1.853 & 1.815 & 1.768 & 1.722 & 1.685 & 1.653 & 1.626 \\ \hline
       \textbf{6} & 1.974 & 1.946 & 1.914 & 1.876 & 1.829 & 1.782 & 1.744 & 1.712 & 1.684 \\ \hline
       \textbf{7} & 2.027 & 1.999 & 1.967 & 1.929 & 1.882 & 1.835 & 1.796 & 1.763 & 1.734 \\ \hline
       \textbf{8} & 2.074 & 2.047 & 2.015 & 1.976 & 1.929 & 1.881 & 1.842 & 1.808 & 1.778 \\ \hline
       \textbf{9} & 2.116 & 2.089 & 2.057 & 2.019 & 1.972 & 1.923 & 1.883 & 1.848 & 1.818 \\ \hline
       \textbf{10} & 2.155 & 2.128 & 2.096 & 2.058 & 2.010 & 1.961 & 1.920 & 1.885 & 1.854 \\ \hline
       \textbf{11} & 2.191 & 2.164 & 2.132 & 2.094 & 2.046 & 1.996 & 1.954 & 1.918 & 1.886 \\ \hline
       \textbf{12} & 2.224 & 2.197 & 2.165 & 2.127 & 2.078 & 2.028 & 1.986 & 1.949 & 1.916 \\ \hline
       \textbf{13} & 2.254 & 2.228 & 2.196 & 2.157 & 2.109 & 2.058 & 2.015 & 1.978 & 1.944 \\ \hline
       \textbf{14} & 2.283 & 2.256 & 2.225 & 2.186 & 2.137 & 2.086 & 2.042 & 2.004 & 1.970 \\ \hline
       \textbf{15} & 2.310 & 2.283 & 2.252 & 2.213 & 2.164 & 2.112 & 2.068 & 2.029 & 1.995 \\ \hline
       \textbf{16} & 2.336 & 2.309 & 2.277 & 2.238 & 2.189 & 2.137 & 2.092 & 2.053 & 2.018 \\ \hline
       \textbf{17} & 2.360 & 2.333 & 2.301 & 2.262 & 2.212 & 2.160 & 2.115 & 2.075 & 2.039 \\ \hline
       \textbf{18} & 2.383 & 2.356 & 2.324 & 2.285 & 2.235 & 2.182 & 2.137 & 2.096 & 2.060 \\ \hline
       \textbf{19} & 2.404 & 2.378 & 2.346 & 2.307 & 2.256 & 2.203 & 2.157 & 2.116 & 2.079 \\ \hline
       \textbf{20} & 2.425 & 2.399 & 2.367 & 2.327 & 2.277 & 2.223 & 2.177 & 2.135 & 2.098 \\ \hline
   \end{tabular}
   \caption{Rényi entropy for the harmonic oscillator  in momentum space vs $n$ for $\omega=1$ and different values of $\pe$. (Data plotted in Figure \ref{grid: momentum p effect} A).}
   \label{p9}
\end{table}

\begin{table}[!ht]
   \centering
   \begin{tabular}{|c|c|c|c|c|c|c|c|c|c|}
   \hline
       \diagbox{\boldsymbol{$n$}}{\boldsymbol{$\alpha$}} & \textbf{0.5} & \textbf{0.57} & \textbf{1} & \textbf{0.67} & \textbf{0.8} & \textbf{1.25} & \textbf{1.5} & \textbf{1.75} & \textbf{2} \\ \hline
       \textbf{0} & 1.004 & 0.961 & 0.916 & 0.868 & 0.814 & 0.764 & 0.726 & 0.695 & 0.670 \\ \hline
       \textbf{1} & 1.077 & 1.036 & 0.993 & 0.947 & 0.898 & 0.853 & 0.819 & 0.792 & 0.770 \\ \hline
       \textbf{2} & 1.083 & 1.041 & 0.997 & 0.950 & 0.899 & 0.853 & 0.817 & 0.788 & 0.764 \\ \hline
       \textbf{3} & 1.073 & 1.031 & 0.987 & 0.940 & 0.887 & 0.839 & 0.801 & 0.770 & 0.744 \\ \hline
       \textbf{4} & 1.061 & 1.019 & 0.975 & 0.927 & 0.874 & 0.824 & 0.785 & 0.753 & 0.725 \\ \hline
       \textbf{5} & 1.049 & 1.008 & 0.964 & 0.917 & 0.863 & 0.813 & 0.773 & 0.740 & 0.712 \\ \hline
       \textbf{6} & 1.039 & 0.998 & 0.955 & 0.908 & 0.854 & 0.804 & 0.764 & 0.731 & 0.703 \\ \hline
       \textbf{7} & 1.030 & 0.990 & 0.947 & 0.900 & 0.847 & 0.797 & 0.757 & 0.724 & 0.696 \\ \hline
       \textbf{8} & 1.022 & 0.983 & 0.941 & 0.894 & 0.841 & 0.791 & 0.752 & 0.719 & 0.691 \\ \hline
       \textbf{9} & 1.015 & 0.977 & 0.935 & 0.889 & 0.837 & 0.787 & 0.748 & 0.715 & 0.687 \\ \hline
       \textbf{10} & 1.009 & 0.971 & 0.931 & 0.885 & 0.833 & 0.783 & 0.744 & 0.712 & 0.684 \\ \hline
       \textbf{11} & 1.004 & 0.967 & 0.926 & 0.882 & 0.830 & 0.781 & 0.742 & 0.709 & 0.682 \\ \hline
       \textbf{12} & 0.999 & 0.963 & 0.923 & 0.878 & 0.827 & 0.778 & 0.739 & 0.707 & 0.680 \\ \hline
       \textbf{13} & 0.995 & 0.959 & 0.920 & 0.876 & 0.825 & 0.776 & 0.738 & 0.706 & 0.679 \\ \hline
       \textbf{14} & 0.991 & 0.956 & 0.917 & 0.873 & 0.823 & 0.775 & 0.736 & 0.704 & 0.677 \\ \hline
       \textbf{15} & 0.988 & 0.953 & 0.914 & 0.871 & 0.821 & 0.773 & 0.735 & 0.703 & 0.676 \\ \hline
       \textbf{16} & 0.985 & 0.950 & 0.912 & 0.869 & 0.820 & 0.772 & 0.734 & 0.702 & 0.676 \\ \hline
       \textbf{17} & 0.982 & 0.948 & 0.910 & 0.868 & 0.818 & 0.771 & 0.733 & 0.702 & 0.675 \\ \hline
       \textbf{18} & 0.980 & 0.946 & 0.908 & 0.866 & 0.817 & 0.770 & 0.732 & 0.701 & 0.674 \\ \hline
       \textbf{19} & 0.977 & 0.944 & 0.907 & 0.865 & 0.816 & 0.769 & 0.731 & 0.700 & 0.674 \\ \hline
       \textbf{20} & 0.975 & 0.942 & 0.905 & 0.864 & 0.815 & 0.768 & 0.731 & 0.700 & 0.673 \\ \hline
   \end{tabular}
   \caption{Rényi entropy for the Darboux III oscillator $(\lambda=0.4)$  in momentum space vs $n$ for $\omega=1$ and different values of $\pe$. (Data plotted in Figure \ref{grid: momentum p effect} B).}
   \label{p10}
\end{table}

\begin{table}[!ht]
   \centering
   \begin{tabular}{|c|c|c|c|c|c|c|c|c|c|}
   \hline
       \diagbox{\boldsymbol{$n$}}{\boldsymbol{$\alpha$}} & \textbf{0.5} & \textbf{0.57} & \textbf{1} & \textbf{0.67} & \textbf{0.8} & \textbf{1.25} & \textbf{1.5} & \textbf{1.75} & \textbf{2} \\ \hline
       \textbf{0} & 1.766 & 1.611 & 1.447 & 1.268 & 1.072 & 0.899 & 0.773 & 0.677 & 0.601 \\ \hline
       \textbf{1} & 2.249 & 2.054 & 1.841 & 1.606 & 1.343 & 1.107 & 0.935 & 0.804 & 0.701 \\ \hline
       \textbf{2} & 2.571 & 2.342 & 2.091 & 1.812 & 1.499 & 1.218 & 1.015 & 0.863 & 0.744 \\ \hline
       \textbf{3} & 2.820 & 2.563 & 2.280 & 1.964 & 1.610 & 1.295 & 1.069 & 0.900 & 0.771 \\ \hline
       \textbf{4} & 3.028 & 2.746 & 2.433 & 2.086 & 1.697 & 1.353 & 1.108 & 0.927 & 0.789 \\ \hline
       \textbf{5} & 3.208 & 2.902 & 2.564 & 2.188 & 1.768 & 1.399 & 1.139 & 0.947 & 0.803 \\ \hline
       \textbf{6} & 3.367 & 3.040 & 2.678 & 2.276 & 1.829 & 1.438 & 1.164 & 0.964 & 0.814 \\ \hline
       \textbf{7} & 3.510 & 3.163 & 2.780 & 2.354 & 1.882 & 1.472 & 1.185 & 0.978 & 0.823 \\ \hline
       \textbf{8} & 3.641 & 3.276 & 2.872 & 2.424 & 1.929 & 1.501 & 1.204 & 0.990 & 0.831 \\ \hline
       \textbf{9} & 3.762 & 3.379 & 2.956 & 2.488 & 1.972 & 1.527 & 1.220 & 1.000 & 0.838 \\ \hline
       \textbf{10} & 3.875 & 3.475 & 3.034 & 2.546 & 2.010 & 1.550 & 1.234 & 1.009 & 0.843 \\ \hline
       \textbf{11} & 3.981 & 3.565 & 3.106 & 2.600 & 2.046 & 1.571 & 1.247 & 1.017 & 0.848 \\ \hline
       \textbf{12} & 4.080 & 3.649 & 3.174 & 2.650 & 2.078 & 1.591 & 1.259 & 1.024 & 0.853 \\ \hline
       \textbf{13} & 4.174 & 3.728 & 3.237 & 2.697 & 2.109 & 1.609 & 1.270 & 1.031 & 0.857 \\ \hline
       \textbf{14} & 4.263 & 3.803 & 3.297 & 2.742 & 2.137 & 1.625 & 1.280 & 1.037 & 0.861 \\ \hline
       \textbf{15} & 4.349 & 3.875 & 3.354 & 2.783 & 2.164 & 1.641 & 1.289 & 1.042 & 0.864 \\ \hline
       \textbf{16} & 4.430 & 3.943 & 3.409 & 2.823 & 2.189 & 1.655 & 1.297 & 1.047 & 0.867 \\ \hline
       \textbf{17} & 4.508 & 4.008 & 3.460 & 2.861 & 2.212 & 1.669 & 1.305 & 1.052 & 0.870 \\ \hline
       \textbf{18} & 4.583 & 4.071 & 3.510 & 2.897 & 2.235 & 1.682 & 1.313 & 1.057 & 0.873 \\ \hline
       \textbf{19} & 4.655 & 4.131 & 3.557 & 2.931 & 2.256 & 1.694 & 1.320 & 1.061 & 0.875 \\ \hline
       \textbf{20} & 4.725 & 4.189 & 3.603 & 2.964 & 2.277 & 1.705 & 1.326 & 1.065 & 0.877 \\ \hline
   \end{tabular}
   \caption{Tsallis entropy for the harmonic oscillator  in momentum space vs $n$ for $\omega=1$ and different values of $\pe$. (Data plotted in Figure \ref{grid: momentum p effect} C).}
   \label{p11}
\end{table}

\begin{table}[!ht]
   \centering
   \begin{tabular}{|c|c|c|c|c|c|c|c|c|c|}
   \hline
       \diagbox{\boldsymbol{$n$}}{\boldsymbol{$\alpha$}} & \textbf{0.5} & \textbf{0.57} & \textbf{1} & \textbf{0.67} & \textbf{0.8} & \textbf{1.25} & \textbf{1.5} & \textbf{1.75} & \textbf{2} \\ \hline
       \textbf{0} & 1.304 & 1.189 & 1.071 & 0.948 & 0.814 & 0.695 & 0.609 & 0.542 & 0.488 \\ \hline
       \textbf{1} & 1.427 & 1.303 & 1.176 & 1.043 & 0.898 & 0.768 & 0.672 & 0.597 & 0.537 \\ \hline
       \textbf{2} & 1.436 & 1.311 & 1.182 & 1.047 & 0.899 & 0.768 & 0.671 & 0.595 & 0.534 \\ \hline
       \textbf{3} & 1.420 & 1.297 & 1.169 & 1.034 & 0.887 & 0.757 & 0.660 & 0.585 & 0.525 \\ \hline
       \textbf{4} & 1.399 & 1.278 & 1.152 & 1.019 & 0.874 & 0.745 & 0.649 & 0.575 & 0.516 \\ \hline
       \textbf{5} & 1.379 & 1.261 & 1.137 & 1.006 & 0.863 & 0.735 & 0.641 & 0.568 & 0.509 \\ \hline
       \textbf{6} & 1.362 & 1.246 & 1.125 & 0.995 & 0.854 & 0.728 & 0.635 & 0.563 & 0.505 \\ \hline
       \textbf{7} & 1.347 & 1.233 & 1.114 & 0.987 & 0.847 & 0.722 & 0.630 & 0.559 & 0.501 \\ \hline
       \textbf{8} & 1.334 & 1.222 & 1.105 & 0.979 & 0.841 & 0.718 & 0.627 & 0.556 & 0.499 \\ \hline
       \textbf{9} & 1.322 & 1.213 & 1.098 & 0.973 & 0.837 & 0.714 & 0.624 & 0.553 & 0.497 \\ \hline
       \textbf{10} & 1.313 & 1.205 & 1.091 & 0.968 & 0.833 & 0.712 & 0.621 & 0.552 & 0.496 \\ \hline
       \textbf{11} & 1.304 & 1.198 & 1.085 & 0.964 & 0.830 & 0.709 & 0.620 & 0.550 & 0.494 \\ \hline
       \textbf{12} & 1.296 & 1.192 & 1.081 & 0.960 & 0.827 & 0.707 & 0.618 & 0.549 & 0.493 \\ \hline
       \textbf{13} & 1.289 & 1.186 & 1.076 & 0.957 & 0.825 & 0.706 & 0.617 & 0.548 & 0.493 \\ \hline
       \textbf{14} & 1.283 & 1.181 & 1.072 & 0.954 & 0.823 & 0.704 & 0.616 & 0.547 & 0.492 \\ \hline
       \textbf{15} & 1.278 & 1.177 & 1.069 & 0.952 & 0.821 & 0.703 & 0.615 & 0.547 & 0.492 \\ \hline
       \textbf{16} & 1.273 & 1.173 & 1.066 & 0.950 & 0.820 & 0.702 & 0.614 & 0.546 & 0.491 \\ \hline
       \textbf{17} & 1.268 & 1.169 & 1.063 & 0.948 & 0.818 & 0.701 & 0.614 & 0.545 & 0.491 \\ \hline
       \textbf{18} & 1.264 & 1.166 & 1.061 & 0.946 & 0.817 & 0.700 & 0.613 & 0.545 & 0.490 \\ \hline
       \textbf{19} & 1.260 & 1.163 & 1.058 & 0.944 & 0.816 & 0.699 & 0.612 & 0.545 & 0.490 \\ \hline
       \textbf{20} & 1.257 & 1.160 & 1.056 & 0.943 & 0.815 & 0.699 & 0.612 & 0.544 & 0.490 \\ \hline
   \end{tabular}
   \caption{Tsallis entropy for the Darboux III oscillator $(\lambda=0.4)$  in momentum space vs $n$ for $\omega=1$ and different values of $\pe$. (Data plotted in Figure \ref{grid: momentum p effect} D).}
   \label{p12}
\end{table}

\begin{table}[!ht]
    \centering
    \begin{tabular}{|c|c|c|c|c|c|c|}
    \hline
    \diagbox{\boldsymbol{$\lambda$}}{} & \boldsymbol{$\mathcal{R}^{(\pe,n=0,\lambda)}$} & \boldsymbol{$\mathcal{R}^{(\pe,n=1,\lambda)}$} & \boldsymbol{$\mathcal{R}^{(\pe,n=2,\lambda)}$} & \boldsymbol{$\mathcal{T}^{(\pe,n=0,\lambda)}$} & \boldsymbol{$\mathcal{T}^{(\pe,n=1,\lambda)}$} & \boldsymbol{$\mathcal{T}^{(\pe,n=2,\lambda)}$} \\ \hline
    \textbf{0,00} & 0,9189 & 1,2066 & 1,3642 & 0,6011 & 0,7008 & 0,7444 \\ \hline
        \textbf{0,05} & 0,8825 & 1,1442 & 1,2836 & 0,5862 & 0,6815 & 0,7230 \\ \hline
        \textbf{0,10} & 0,8478 & 1,0829 & 1,2000 & 0,5717 & 0,6614 & 0,6988 \\ \hline
        \textbf{0,15} & 0,8149 & 1,0235 & 1,1169 & 0,5573 & 0,6407 & 0,6727 \\ \hline
        \textbf{0,20} & 0,7834 & 0,9666 & 1,0367 & 0,5431 & 0,6196 & 0,6454 \\ \hline
        \textbf{0,25} & 0,7532 & 0,9128 & 0,9608 & 0,5292 & 0,5986 & 0,6174 \\ \hline
        \textbf{0,30} & 0,7244 & 0,8620 & 0,8900 & 0,5154 & 0,5777 & 0,5893 \\ \hline
        \textbf{0,35} & 0,6968 & 0,8143 & 0,8243 & 0,5018 & 0,5571 & 0,5615 \\ \hline
        \textbf{0,40} & 0,6704 & 0,7697 & 0,7636 & 0,4885 & 0,5368 & 0,5340 \\ \hline
        \textbf{0,45} & 0,6450 & 0,7280 & 0,7076 & 0,4753 & 0,5171 & 0,5072 \\ \hline
        \textbf{0,50} & 0,6207 & 0,6889 & 0,6558 & 0,4624 & 0,4979 & 0,4810 \\ \hline
        \textbf{0,55} & 0,5974 & 0,6523 & 0,6078 & 0,4498 & 0,4792 & 0,4554 \\ \hline
        \textbf{0,60} & 0,5751 & 0,6180 & 0,5632 & 0,4374 & 0,4610 & 0,4306 \\ \hline
        \textbf{0,65} & 0,5537 & 0,5857 & 0,5217 & 0,4252 & 0,4433 & 0,4065 \\ \hline
        \textbf{0,70} & 0,5332 & 0,5552 & 0,4829 & 0,4133 & 0,4260 & 0,3830 \\ \hline
        \textbf{0,75} & 0,5135 & 0,5264 & 0,4465 & 0,4016 & 0,4093 & 0,3601 \\ \hline
        \textbf{0,80} & 0,4946 & 0,4991 & 0,4123 & 0,3902 & 0,3930 & 0,3379 \\ \hline
        \textbf{0,85} & 0,4765 & 0,4732 & 0,3801 & 0,3791 & 0,3770 & 0,3162 \\ \hline
        \textbf{0,90} & 0,4591 & 0,4486 & 0,3497 & 0,3682 & 0,3615 & 0,2951 \\ \hline
        \textbf{0,95} & 0,4424 & 0,4251 & 0,3209 & 0,3575 & 0,3463 & 0,2745 \\ \hline
        \textbf{1,00} & 0,4264 & 0,4026 & 0,2936 & 0,3471 & 0,3314 & 0,2544 \\ \hline
        \textbf{1,05} & 0,4110 & 0,3810 & 0,2676 & 0,3370 & 0,3168 & 0,2348 \\ \hline
        \textbf{1,10} & 0,3962 & 0,3604 & 0,2429 & 0,3271 & 0,3026 & 0,2156 \\ \hline
        \textbf{1,15} & 0,3819 & 0,3405 & 0,2192 & 0,3175 & 0,2886 & 0,1969 \\ \hline
        \textbf{1,20} & 0,3682 & 0,3214 & 0,1966 & 0,3080 & 0,2749 & 0,1785 \\ \hline
        \textbf{1,25} & 0,3550 & 0,3030 & 0,1749 & 0,2988 & 0,2614 & 0,1605 \\ \hline
        \textbf{1,30} & 0,3423 & 0,2852 & 0,1541 & 0,2899 & 0,2481 & 0,1428 \\ \hline
        \textbf{1,35} & 0,3300 & 0,2680 & 0,1341 & 0,2811 & 0,2351 & 0,1255 \\ \hline
        \textbf{1,40} & 0,3182 & 0,2513 & 0,1149 & 0,2725 & 0,2222 & 0,1086 \\ \hline
        \textbf{1,45} & 0,3068 & 0,2352 & 0,0964 & 0,2642 & 0,2096 & 0,0919 \\ \hline
        \textbf{1,50} & 0,2957 & 0,2196 & 0,0785 & 0,2560 & 0,1972 & 0,0755 \\ \hline
    \end{tabular}
    \caption{Rényi and Tsallis entropies in momentum space vs $\lambda$ for $\omega=1$, $n=0,1,2$, $\pe=2$ and $\omega=1$. (Data plotted in Figure \ref{grid: momentum lambda effect}).}
    \label{table: momentum lambda effect}
\end{table}

\begin{table}[!ht]
   \centering
   \begin{tabular}{|c|c|c|c|c|c|c|c|c|}
   \hline
   \diagbox{\boldsymbol{$n$}}{\boldsymbol{$\alpha$}} & \textbf{0.6} & \textbf{0.7} & \textbf{0.8} & \textbf{0.9} & \textbf{1.125} & \textbf{1.333} & \textbf{1.75} & \textbf{3} \\ \hline
       \textbf{0} & 0 & 0 & 0 & 0 & 0 & 0 & 0 & 0 \\ \hline
       \textbf{1} & 0.544 & 0.542 & 0.541 & 0.541 & 0.541 & 0.541 & 0.542 & 0.544 \\ \hline
       \textbf{2} & 0.849 & 0.852 & 0.852 & 0.852 & 0.852 & 0.852 & 0.852 & 0.849 \\ \hline
       \textbf{3} & 1.062 & 1.072 & 1.074 & 1.075 & 1.075 & 1.074 & 1.072 & 1.062 \\ \hline
       \textbf{4} & 1.228 & 1.243 & 1.247 & 1.248 & 1.248 & 1.247 & 1.243 & 1.228 \\ \hline
       \textbf{5} & 1.363 & 1.384 & 1.389 & 1.391 & 1.391 & 1.389 & 1.384 & 1.363 \\ \hline
   \end{tabular}
   \caption{$\xi$ function for the Rényi entropy of the harmonic oscillator vs $n$ for $\omega=1$ and different values of $\pe$. (Data plotted in Figure \ref{grid: uncertainty} A).}
   \label{p17}
\end{table}

\begin{table}[!ht]
   \centering
   \begin{tabular}{|c|c|c|c|c|c|c|c|c|}
   \hline
   \diagbox{\boldsymbol{$n$}}{\boldsymbol{$\alpha$}} & \textbf{0.6} & \textbf{0.7} & \textbf{0.8} & \textbf{0.9} & \textbf{1.125} & \textbf{1.333} & \textbf{1.75} & \textbf{3} \\ \hline
       \textbf{0} & 0.001 & 0.001 & 0.002 & 0.003 & 0.006 & 0.009 & 0.015 & 0.027 \\ \hline
       \textbf{1} & 0.486 & 0.48 & 0.477 & 0.475 & 0.474 & 0.475 & 0.477 & 0.484 \\ \hline
       \textbf{2} & 0.69 & 0.689 & 0.681 & 0.674 & 0.66 & 0.65 & 0.636 & 0.615 \\ \hline
       \textbf{3} & 0.825 & 0.826 & 0.815 & 0.803 & 0.78 & 0.762 & 0.736 & 0.699 \\ \hline
       \textbf{4} & 0.937 & 0.937 & 0.924 & 0.909 & 0.879 & 0.856 & 0.823 & 0.776 \\ \hline
       \textbf{5} & 1.039 & 1.037 & 1.021 & 1.004 & 0.969 & 0.943 & 0.905 & 0.85 \\ \hline
   \end{tabular}
   \caption{$\xi$ function for the Rényi entropy of the Darboux III oscillator $(\lambda=0.4)$ vs $n$ for $\omega=1$ and different values of $\pe$. (Data plotted in Figure \ref{grid: uncertainty} B).}
   \label{p18}
\end{table}

\begin{table}[!ht]
   \centering
   \begin{tabular}{|c|c|c|c|c|}
   \hline
       \diagbox{\boldsymbol{$n$}}{\boldsymbol{$\alpha$}} & \textbf{0.6} & \textbf{0.7} & \textbf{0.8} & \textbf{0.9} \\ \hline
       \textbf{0} & 0 & 0 & 0 & 0 \\ \hline
       \textbf{1} & 0.1350 & 0.0870 & 0.0510 & 0.0230 \\ \hline
       \textbf{2} & 0.2110 & 0.1370 & 0.0800 & 0.0360 \\ \hline
       \textbf{3} & 0.2650 & 0.1720 & 0.1010 & 0.0450 \\ \hline
       \textbf{4} & 0.3070 & 0.2000 & 0.1170 & 0.0520 \\ \hline
       \textbf{5} & 0.342 & 0.223 & 0.13 & 0.058 \\ \hline
   \end{tabular}
   \caption{$\xi$ uncertainty function for the Tsallis entropy of the harmonic oscillator vs $n$ for $\omega=1$ and different values of $\pe$. (Data plotted in Figure \ref{grid: uncertainty} C).}
   \label{p19}
\end{table}

\begin{table}[!ht]
   \centering
   \begin{tabular}{|c|c|c|c|c|}
   \hline
       \diagbox{\boldsymbol{$n$}}{\boldsymbol{$\alpha$}} & \textbf{0.6} & \textbf{0.7} & \textbf{0.8} & \textbf{0.9} \\ \hline
       \textbf{0} & 0.00018 & 0.00024 & 0.00022 & 0.00014 \\ \hline
       \textbf{1} & 0.1380 & 0.0840 & 0.0470 & 0.0200 \\ \hline
       \textbf{2} & 0.2050 & 0.1240 & 0.0680 & 0.0290 \\ \hline
       \textbf{3} & 0.2530 & 0.1510 & 0.0820 & 0.0350 \\ \hline
       \textbf{4} & 0.2950 & 0.1740 & 0.0940 & 0.0390 \\ \hline
       \textbf{5} & 0.334 & 0.195 & 0.105 & 0.044 \\ \hline
   \end{tabular}
   \caption{$\xi$ uncertainty function for the Tsallis entropy of the Darboux III oscillator $(\lambda=0.4)$ vs $n$ for $\omega=1$ and different values of $\pe$. (Data plotted in Figure \ref{grid: uncertainty} D).}
   \label{p20}
\end{table}

\begin{table}[!ht]
    \centering
    \begin{tabular}{|c|c|c|c|}
    \hline
       \diagbox{\boldsymbol{$\lambda$}}{n} & \textbf{0} & \textbf{1} & \textbf{2} \\ \hline
        \textbf{0} & 0 & 0,54273364 & 0,85144604 \\ \hline
        \textbf{0,1} & 0,00003950 & 0,53361714 & 0,80968346 \\ \hline
        \textbf{0,2} & 0,00121591 & 0,51289017 & 0,73743021 \\ \hline
        \textbf{0,3} & 0,00665309 & 0,49307262 & 0,67406665 \\ \hline
        \textbf{0,4} & 0,01796345 & 0,47909635 & 0,62979980 \\ \hline
        \textbf{0,5} & 0,03466344 & 0,47055134 & 0,60106409 \\ \hline
        \textbf{0,6} & 0,05542587 & 0,46594978 & 0,58279272 \\ \hline
        \textbf{0,7} & 0,07880639 & 0,46398802 & 0,57121822 \\ \hline
        \textbf{0,8} & 0,10353287 & 0,46372208 & 0,56389772 \\ \hline
        \textbf{0,9} & 0,12859443 & 0,46450774 & 0,55930815 \\ \hline
        \textbf{1} & 0,15324539 & 0,46591630 & 0,55649525 \\ \hline
        \textbf{1,1} & 0,17697464 & 0,46766686 & 0,55485328 \\ \hline
        \textbf{1,2} & 0,19946287 & 0,46957876 & 0,55399687 \\ \hline
        \textbf{1,3} & 0,22053870 & 0,47153762 & 0,55366900 \\ \hline
        \textbf{1,4} & 0,24013894 & 0,47347255 & 0,55370207 \\ \hline
        \textbf{1,5} & 0,25827613 & 0,47534141 & 0,55397462 \\ \hline
        \textbf{1,6} & 0,27501066 & 0,47712036 & 0,55441554 \\ \hline
        \textbf{1,7} & 0,29043252 & 0,47879774 & 0,55496722 \\ \hline
        \textbf{1,8} & 0,30464556 & 0,48036911 & 0,55558765 \\ \hline
        \textbf{1,9} & 0,31775899 & 0,48183592 & 0,55625299 \\ \hline
        \textbf{2} & 0,32987899 & 0,48320054 & 0,55693940 \\ \hline
        \textbf{2,1} & 0,34110731 & 0,48446892 & 0,55763184 \\ \hline
        \textbf{2,2} & 0,35153641 & 0,48564663 & 0,55833342 \\ \hline
        \textbf{2,3} & 0,36125079 & 0,48673929 & 0,55901693 \\ \hline
        \textbf{2,4} & 0,37032594 & 0,48775468 & 0,55969952 \\ \hline
        \textbf{2,5} & 0,37882788 & 0,48869707 & 0,56036331 \\ \hline
        \textbf{2,6} & 0,38681610 & 0,48957338 & 0,56100858 \\ \hline
        \textbf{2,7} & 0,39434204 & 0,49038818 & 0,56163327 \\ \hline
        \textbf{2,8} & 0,40145013 & 0,49114692 & 0,56223514 \\ \hline
        \textbf{2,9} & 0,40818008 & 0,49185330 & 0,56282318 \\ \hline
        \textbf{3} & 0,41456658 & 0,49251278 & 0,56339437 \\ \hline
    \end{tabular}
    \caption{$\xi$ uncertainty function of the Rényi entropy vs $\lambda$ for $\omega=1$ and $n=0,1,2$. (Data plotted in Figure \ref{grid: uncertainty vs lambda} A).}
    \label{table: uncertainty vs lambda A}
\end{table}

\begin{table}[!ht]
    \centering
    \begin{tabular}{|c|c|c|c|}
    \hline
       \diagbox{\boldsymbol{$\lambda$}}{n} & \textbf{0} & \textbf{1} & \textbf{2} \\ \hline
        \textbf{0} & 0 & 0,00298405 & 0,00463809 \\ \hline
        \textbf{0,1} & 0,00008063 & 0,00303296 & 0,00474984 \\ \hline
        \textbf{0,2} & 0,00008264 & 0,00302369 & 0,00466016 \\ \hline
        \textbf{0,3} & 0,00008584 & 0,00298681 & 0,00448607 \\ \hline
        \textbf{0,4} & 0,00009065 & 0,00294090 & 0,00429998 \\ \hline
        \textbf{0,5} & 0,00009729 & 0,00289627 & 0,00413297 \\ \hline
        \textbf{0,6} & 0,00010583 & 0,00285766 & 0,00399344 \\ \hline
        \textbf{0,7} & 0,00011623 & 0,00282657 & 0,00388047 \\ \hline
        \textbf{0,8} & 0,00012838 & 0,00280284 & 0,00379024 \\ \hline
        \textbf{0,9} & 0,00014212 & 0,00278561 & 0,00371859 \\ \hline
        \textbf{1} & 0,00015724 & 0,00277385 & 0,00366183 \\ \hline
        \textbf{1,1} & 0,00017354 & 0,00276654 & 0,00361693 \\ \hline
        \textbf{1,2} & 0,00019082 & 0,00276281 & 0,00358149 \\ \hline
        \textbf{1,3} & 0,00020886 & 0,00276191 & 0,00355363 \\ \hline
        \textbf{1,4} & 0,00022749 & 0,00276325 & 0,00353186 \\ \hline
        \textbf{1,5} & 0,00024652 & 0,00276636 & 0,00351501 \\ \hline
        \textbf{1,6} & 0,00026581 & 0,00277085 & 0,00350218 \\ \hline
        \textbf{1,7} & 0,00028522 & 0,00277641 & 0,00349263 \\ \hline
        \textbf{1,8} & 0,00030464 & 0,00278281 & 0,00348578 \\ \hline
        \textbf{1,9} & 0,00032396 & 0,00278986 & 0,00348115 \\ \hline
        \textbf{2} & 0,00034311 & 0,00279740 & 0,00347838 \\ \hline
        \textbf{2,1} & 0,00036202 & 0,00280531 & 0,00347715 \\ \hline
        \textbf{2,2} & 0,00038063 & 0,00281350 & 0,00347721 \\ \hline
        \textbf{2,3} & 0,00039890 & 0,00282189 & 0,00347835 \\ \hline
        \textbf{2,4} & 0,00041681 & 0,00283041 & 0,00348040 \\ \hline
        \textbf{2,5} & 0,00043432 & 0,00283903 & 0,00348321 \\ \hline
        \textbf{2,6} & 0,00045141 & 0,00284769 & 0,00348667 \\ \hline
        \textbf{2,7} & 0,00046809 & 0,00285636 & 0,00349067 \\ \hline
        \textbf{2,8} & 0,00048433 & 0,00286503 & 0,00349513 \\ \hline
        \textbf{2,9} & 0,00050015 & 0,00287367 & 0,00349998 \\ \hline
        \textbf{3} & 0,00051553 & 0,00288225 & 0,00350516 \\ \hline
    \end{tabular}
    \caption{$\xi$ uncertainty function of the Rényi entropy vs $\lambda$ for $\omega=1$ and $n=0,1,2$. (Data plotted in Figure \ref{grid: uncertainty vs lambda} B).}
    \label{table: uncertainty vs lambda B}
\end{table}

\end{document}